\documentclass[10pt, conference, letterpaper]{IEEEtran}
\usepackage{array}
\usepackage{amsmath, amsthm, amssymb, amsfonts}
\usepackage{algorithm,algorithmicx}
\usepackage[noend]{algpseudocode}
\usepackage[export]{adjustbox}
\usepackage{balance}
\usepackage{bm}
\usepackage{booktabs, tabularx}
\usepackage[skip=4pt,font={bf}]{caption}  %shrink...
\usepackage{enumitem}
\usepackage{float}
\usepackage{graphicx}
\usepackage{subfigure}
\usepackage[noadjust]{cite}
\usepackage{multirow, multicol, hhline}
\usepackage{siunitx}
\usepackage{tikz}
\usepackage{tikzpeople}
\usepackage{thmtools, thm-restate}
\usepackage{color, xcolor}

\allowdisplaybreaks[1]
\DeclareGraphicsExtensions{.pdf,.png,.jpg}

\ifx\papertech\undefined
  \ifx\techreport\undefined
    \newcommand{\onlytech}[1]{\ignorespaces}
    \newcommand{\onlypaper}[1]{{\color{black}#1}}
  \else
    \newcommand{\onlytech}[1]{#1}
    \newcommand{\onlypaper}[1]{\ignorespaces}
    \makeatletter
    \def\@copyrightspace{\relax}
    \makeatother
  \fi
\else
  \newcommand{\onlytech}[1]{{\color{blue}[(OLD) #1]}}
  \newcommand{\onlypaper}[1]{{\color{black}[(REVISED) #1]}}
\fi

\makeatletter%
\@ifclassloaded{acmart}
{}{
  \usepackage{substitutefont, newtxtext} % texttt
}
\makeatother

\usepackage{url}
\definecolor{red}{HTML}{E51400}  %red
\definecolor{blue}{HTML}{0050EF} %cobalt
\definecolor{green}{HTML}{008A00} %emerald
\definecolor{purple}{HTML}{AA00FF} %violet
\definecolor{dark-red}{rgb}{0.4, 0.15, 0.15}
\definecolor{dark-blue}{rgb}{0.15, 0.15, 0.4}
\definecolor{medium-red}{rgb}{0.5, 0, 0}
\definecolor{medium-blue}{rgb}{0, 0, 0.5}
\definecolor{light-red}{rgb}{0.7, 0, 0}
\definecolor{light-blue}{rgb}{0, 0, 0.7}
\definecolor{myOrg}{rgb}{1.0, 0.8, 0.0}
\definecolor{myGrn}{RGB}{34, 139, 34}
\definecolor{myBlue}{RGB}{30, 144, 255}

\pgfdeclarelayer{background}
\pgfdeclarelayer{foreground}
\pgfsetlayers{background,main,foreground}

\newtheoremstyle{exampstyle}% Name of the style to be used
  {0.15\topsep} % Measure of space to leave above the theorem. E.g.: 3pt
  {0.15\topsep} % Measure of space to leave below the theorem. E.g.: 3pt
  {\slshape} 			% Name of font to use in the body of the theorem. E.g.: \slshape
  {} 			% Measure of space to indent. E.g.: 3pt
  {\bfseries} 	% Name of head font. E.g.: \bfseries
  {.} 			% punctuation between head and body
  {.5em} 		% space after theorem head
  {} 			% manually specify theorem head spec (can be left empty, meaning `normal')
\theoremstyle{exampstyle}\newtheorem{theorem}{\hspace{0in}{\bf Theorem}}
\newtheorem{lemman}{\hspace{0in}{\bf Lemma}}

\newtheorem{definition}{\hspace{0in}{\bf Definition}}
\def\done{\hspace*{\fill}\rule{1.8mm}{2.5mm}}



\makeatletter
\def\footnoterule{\kern-3\p@
  \hrule \@width 2in \kern 2.6\p@} % the \hrule is .4pt high
\makeatother 

\newcolumntype{C}[1]{>{\centering\arraybackslash}m{#1}}
\newcommand{\SC}[4]{\multicolumn{1}{#3>{\centering\arraybackslash}m{#1}#4}{#2}}

\newcommand{\compilehidecomments}{false}
\ifthenelse{\equal{\compilehidecomments}{true}}{%
  \newcommand{\vicky}[1]{}
}{
  \newcommand{\vicky}[1]{{\color{red}[\text{Vicky:} #1]}}
}

\newcommand{\mk}[1]{{\color{black}#1}}

\usetikzlibrary{shapes.geometric}%
\usetikzlibrary{positioning}%
\usetikzlibrary{arrows}%
\usetikzlibrary{fit}%
\usetikzlibrary{decorations.markings}
\usetikzlibrary{decorations.shapes}
\usetikzlibrary{calc}
\usetikzlibrary{backgrounds}

\newcommand{\BK}[1]{[{\kern-0.70pt#1\kern-0.70pt}]} 
\newcommand{\bk}[1]{({\kern-0.70pt#1\kern-0.70pt})}

\title{FAVE:~A~fast~and~efficient~network~Flow~AVailability Estimation method with bounded relative error
\thanks{This work is supported in part by the GRF 14200117 and the Huawei Grant.}}
\author{
\IEEEauthorblockN{
	Tingwei Liu$^{\dagger}$,
        John C.S. Lui$^{\ast}$,}
\IEEEauthorblockA{
	\textit{
		Department of Computer Science \& Engineering, 
		The Chinese University of Hong Kong, China} \\
	Email: $^{\dagger}$twliu@cse.cuhk.edu.hk, 
		  $^{\ast}$cslui@cse.cuhk.edu.hk}
}
\IEEEoverridecommandlockouts
\begin{document}
\bstctlcite{BSTcontrol}
\maketitle

\begin{abstract}
This paper focuses on helping network providers~to carry out network capacity planning and sales projection~by~ans-wering the question: For a given topology and capacity,~whether the network can serve current flow demands with high probabili-ties? We name such probability as ``{\it flow availability}", and present the~\underline{f}low \underline{av}ailability \underline{e}stimation (FAVE) problem, which is a gener-alisation of network connectivity or maximum flow reliability~esti-mations.
Realistic networks are~often large and dynamic, so flow availabilities cannot be evaluated analytically and simulation is often used. However, naive Monte Carlo (MC) or importance~sam-pling (IS) techniques take an excessive amount of time. To quickly\newline estimate flow availabilities, we utilize the correlations among link and flow failures to figure out the importance of roles played~by different links in flow failures\onlytech{ (i.e., flow demands are not satisfied)}, and design three ``sequential~imp-ortance sampling" (SIS) methods which achieve ``bounded or even vanishing relative error" with linear computational complexities.\newline When applying to a realistic network, our method reduces~the~flow\!\newline availability estimation cost by 900 and 130 times compared with MC and baseline IS methods, respectively. Our method can also facilitate capacity planning by providing better flow availability guarantees, compared with traditional methods.
\end{abstract}

%\begin{IEEEkeywords}
%Flow availability, sequential importance sampling (SIS), 
%bounded relative error (BRE),
%vanishing relative error (VRE)
%\end{IEEEkeywords}

\setlength{\abovedisplayskip}{2pt}
\setlength{\belowdisplayskip}{2pt}

%%%
\section{\textbf{Introduction}}\label{sec:intro}
Network capacity planning is the process of
ensuring~sufficient bandwidth is provisioned 
so that service-level agreement (SLA) objectives
like delay, jitter, loss and routing~availability
can be satisfied \cite{Cisco-MATE2}.
For the purpose of providing a better~end user experience
and at the same time, keeping the operation cost at an affordable level,
effective capacity planning~tools~are crucial to network providers.
Various systems have been built around this problem,
such as Cisco's MATE \cite{Cisco-MATE},
Facebook's Prophet \cite{Facebook-Prophet},
and Google's backbone \onlytech{network} capacity planning tool \cite{Google-WAND}.

\begin{figure}[tp]\centering
\includegraphics[width=0.45\textwidth,height=0.145\textwidth]{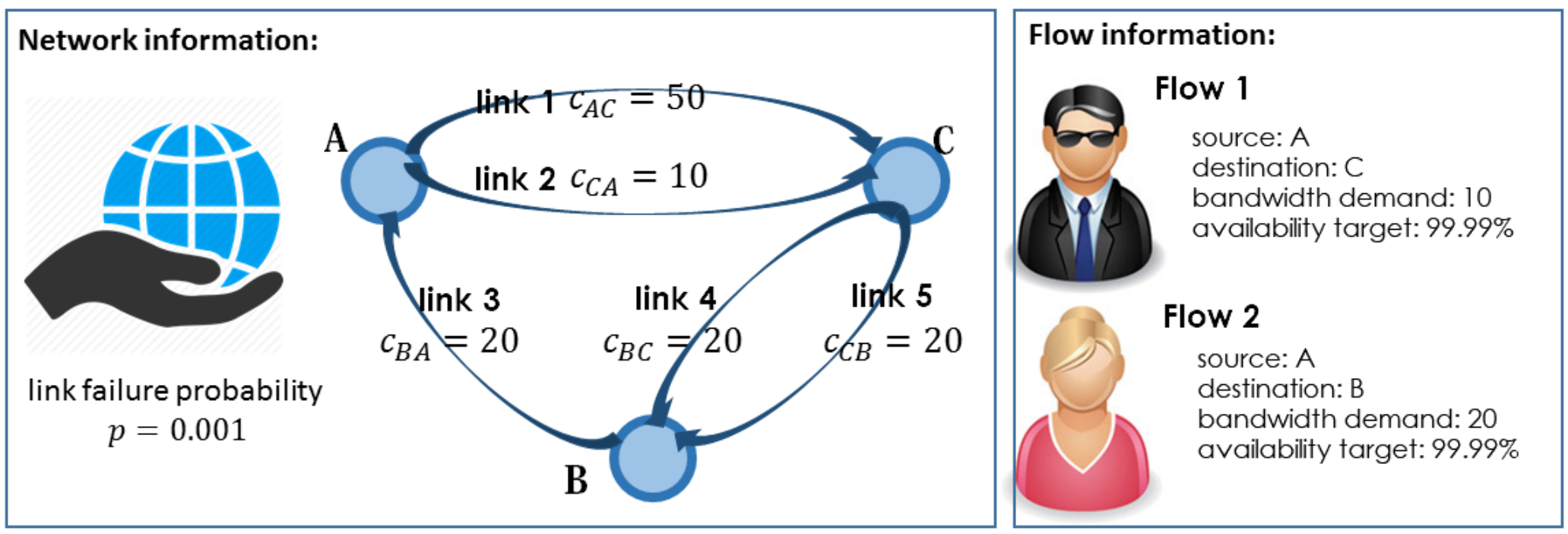}
\caption{An example of the FAVE problem}.\label{fig:intro_eg}
\vspace{-2.5pt}
\end{figure}

\vspace{1pt}
\noindent{\bf Motivations of fast flow availability estimations}:
The above SLA objectives are called the {\it demand} of a traffic flow,
and~the\newline satisfaction probability of demand
is defined as the {\it flow~availability}.
A key concern of capacity planning is in analyzing~the
effect of network changes or the arising of new flows
on~the flow demand satisfaction~\cite{Cisco-MATE}.
%To illustrate, we start with~an~example that a network provider
To illustrate, consider an~example that a network provider
needs to serve two flows as shown in Fig.\ref{fig:intro_eg}.
In this network, each link $i$ is associated with a failure probability,
say $p{=}0.001$ and a capacity~$c_i$;
also, flow routing follows the ``max-min fairness" and
``shortest~path" policies.
Each flow has a bandwidth demand, and an {\it availability target}
(i.e., the lower bound probability that
its~requested bandwidth needs to be satisfied).
The network provider will perform:
1) {\bf Flow availability testing}, 
i.e., whether flow availability~targets can be achieved? 
E.g., flow 1's availability target is achieved if 
it obtains 10 units of bandwidth with a probability no less than $0.9999$.
2) {\bf Capacity planning},
e.g, to improve the network, 
should the provider add more links between node $A$ and $B$,
or add a new node $D$ so to increase the path~diversity?
3) {\bf Sales\newline projection},
e.g., to increase profit, 
can the provider admit~a~new\!\!\newline flow 3?
Will the network still provide flow availability guarantees 
if flow 1 requires 20 additional units of bandwidth?
All these cases need flow availabilities to see whether
the~network\newline (proposed by capacity planning)
can serve flow demands~(proposed by sales projection).
We name the flow availability~esti-mation problem as ``FAVE",
and give a formal~definition~later.

\begin{table*}[!t]
\footnotesize\centering
\caption{Accuracy, efficiency and computational cost (i.e., simulation steps) comparison}\label{tb:intro_eg}
\resizebox{1.8\columnwidth}{0.10\columnwidth}{
\begin{tabular*}{1.90\columnwidth}{|p{0.75cm}||p{1.9cm}||p{1.9cm}p{1.9cm}p{1.45cm}||p{1.9cm}p{1.9cm}p{1.45cm}|}
  \hline
      &  &\multicolumn{3}{c||}{{\bf Monte Carlo Method}} &\multicolumn{3}{c|}{{\bf Our method (SEED-VRE)}}\\
  {\bf Flow}&\SC{1.9cm}{Theoretical\newline Unavailability}{}{||}
  &\SC{1.9cm}{Estimated\newline Unavailability}{}{} &\SC{1.9cm}{Empirical\newline Variance}{}{} &\SC{1.45cm}{Number of\newline Simulations}{}{||}
  &\SC{1.9cm}{Estimated\newline Unavailability}{}{} &\SC{1.9cm}{Empirical\newline Variance}{}{} &\SC{1.45cm}{Number of\newline Simulations}{}{|}\\
  \hline
  \quad 1       &$1.000999{\times}10^{-3}$
                     &$1.300{\times}10^{-3}$                      &$1.29831{\times}10^{-7}$   &\quad10,000
                     &$1.000999{\times}10^{-3}$                &$6.35275{\times}10^{-23}$ &\qquad10  \\
  \quad 2       &$1.000000{\times}10^{-3}$
                     &$8.000000{\times}10^{-4}$                &$7.99360{\times}10^{-8}$   &\quad10,000
                     &$1.000000{\times}10^{-3}$                &$2.11758{\times}10^{-23}$ &\qquad10 \\
  \hline
\end{tabular*}}
\vspace{3pt}
\end{table*}

\vspace{-1pt}
For a large realistic network with complex failure patterns,\newline
the flow availability often cannot be evaluated analytically,~especially
when the traffic scheduling is also considered.
Hence, simulation (or sampling) is often used.
The Monte~Carlo~(MC)\newline method~\cite{hammersley1964-MC_def},
which simulates link failures with real link failure\newline~ probabilities, 
is the most widely used. 
{\color{black}
%Yet,~to achieve a desired accuracy level, 
%the simulation cost of MC can be very large,
Yet, it can be costly~for MC to achieve a desired accuracy level, 
e.g., to estimate flow 2's availability in Fig.\ref{fig:intro_eg},
by variances collected in Table\! \ref{tb:intro_eg},
MC needs at least 3,840 simulation steps to 
guarantee the 95\% confidence interval (CI) width below $10^{-3}$.}
To simulate a large network with many flows, it is
crucial to find ways to speed~up the simulation process.

\noindent{\bf High level ideas of our work:}
The flow availability estimation\newline
can be more efficient if flow failures
(i.e., flow demands~could not be satisfied) 
happen more frequently by 
{\color{black} taking a proper~distribution to simulate link failures},
which is the key idea~of~``{\it importance sampling} (IS)" \cite{Glynn1989-IS_def}. 
We first introduce a baseline IS solution for FAVE using
``{\it the correlation between link failures\newline and flow failures}" 
to decide ``important links" whose failures are more likely to result in flow failures. 
We then simulate~such links' failures more often 
to speed up simulating flow~failures.
To further improve upon the baseline IS,
we show that some link sets' failures happen more often,
and play a role of ``root causes" of flow failures.
We name such link set as ``SEED".~We\newline utilize 
``{\it the correlation among link failures}" captured by~SEEDs\!\!\newline
to decide ``important link sets"
whose failures are more likely\newline to result in flow failures.
We propose three advanced ``{\it sequential importance sampling (SIS)}" algorithms,
among which the SEED-VRE algorithm has
the ``{\it vanishing relative error (VRE)}" and ``{\it linear computational complexity}".
Table\! \ref{tb:intro_eg} illustrates~the computational reduction of our methods:
Using only {\bf 10 simulation steps},
both empirical variance and estimation error are
much smaller than that of MC with {\bf 10,000 simulation steps}.
{\color{black} In other words, not only we can speed up simulations,
but~{\it the network operator is able to %{experiment more configurations~so to}
achieve a higher accuracy of flow availability estimation
without incurring an additional cost}}.

We emphasize that flow availability estimation can also~faci-litate capacity planning: for flows with unachievable availability targets, we show that allocating more capacities for~the~important links determined by our method would bring~a~better\newline flow availability improvement, compared with traditional~me-thods to allocate more capacities~for links with high \onlytech{bandwidth} utilizations.\!\! 
\onlytech{Applying our method on sales projection~and~top-\newline ology design are also presented in our technical report \cite{tech}.}

\vspace{1.5pt}
\noindent {\bf Contributions:} Our key contribution is in providing efficient and accurate solutions for FAVE. Contributions include:
\begin{itemize}[leftmargin=3mm]\vspace{-1pt}
\item We formally define the network Flow AVailability Estimation\!\newline (FAVE) problem, and show it generalizes the classical network reliability estimation problem~\cite{jiang2012-reliability_mc,Yeh2010-PSO_MC,Ecuyer2011-ZV_SIS,Lee2010-crosslayer_reliability,botev2014reliability,gertsbakh2014permutational,datta2016reliability}.
\item \mk{For FAVE with a single flow, we introduce a novel concept~of\newline ``SEED" and propose three advanced SIS algorithms which have~attractive properties of {\it bounded relative error (BRE)}~or\newline {\it VRE} with {\it linear computational complexities}.}
\item \mk{For FAVE with multiple flows, we propose a mixture SIS method which {\it maintains BRE property and linear computa-tional complexity} when estimating all flows' availabilities.}
\item \mk{For FAVE with a partial SEED set information, we maintain the BRE and VRE properties of flow availability estimations.}
\item Extensive results show that our methods greatly speed~up~the flow availability estimation: 1) Given \mk{ an illustrative network} and full SEED sets or partial SEED sets with good coverage, SEED-VRE achieves {\it variance reductions of 360,000 and 7,000 times} in the single flow case, and {\it 200,000 and~7~times}\newline for around 80\% flows in the multiple flow case,~compared with MC and baseline IS. 2) Given a realistic network and partial SEED sets with poor coverage, SEED-VRE~{\it reduces simulation cost by 900 and 130 times} for around 80\% flows in the multiple flow case, compared with MC and baseline~IS.\!
\item We demonstrate our methods facilitate capacity planning~by
providing more accurate network reliability evaluations
compared with classical methods, and greater flow availability improvements compared with solely using the link capacity utilization information.
\end{itemize}

{\noindent \bf Organizations:} \onlytech{The remainder is organized as follows.}
Section \ref{sec:related-work} introduces related work and some preliminaries.
Section \ref{sec:pro_def_fave} formally defines the FAVE problem.
In Section \ref{sec:alg_design}, we propose~a baseline IS solution for FAVE~and\newline show its error bounds,~then we introduce ``SEED" and present SEED based SIS solutions with better error bounds and linear\newline
computational complexities.
Section \ref{sec:sis-prac} considers more~practi-cal issues, e.g., the multiple flows case and partial SEED sets information case.
In Section \ref{sec:simulation}, we evaluate our methods~on both an illustrative network and a realistic network, i.e., the Abilene network \cite{Abilene}.
Section \ref{sec:app} shows the utility of our methods in capacity planning.
Finally, Section \ref{sec:conclusion} concludes.

\section{\textbf{Related Work \& Preliminaries}}\label{sec:related-work}
Our work can be viewed as a generalisation of previous~work\newline
in estimating {\it network reliability}~\cite{jiang2012-reliability_mc,Yeh2010-PSO_MC,Ecuyer2011-ZV_SIS,Lee2010-crosslayer_reliability,botev2014reliability,gertsbakh2014permutational,datta2016reliability} and is related with {\it importance sampling}~\cite{Glynn1989-IS_def}. Here, we briefly review previous~rel-evant studies and compare them with our work. We also~descr-ibe some concepts so readers can gain a better understanding of sampling methods for the network reliability estimation.\vspace{-5pt}

%%%%%%%%%%%%%%%%%%%%%%%%%%%%%%%%%%%%%%%%%%%
\subsection{\textbf{Network Reliability Estimation}}\label{subsec:net_re}
The most relevant literatures to our work focus on 
evaluating {\it network reliability}. 
To design reliable networks, 
one needs to measure the impact of network failures (e.g., link failures)~on\newline
network performances~\cite{xiong2005novel}.
It is known that the~exact computation of network reliability is \#P-complete,
and computational\newline complexities of all known algorithms are
exponential increasing with the graph scale~\cite{ball1986-computational},
which makes the problem~intractable even for medium sized networks.
Hence, most~work on network reliability evaluation considers sampling methods
to provide reliability estimations, and they can be classified~into\!\newline
``{\it network connectivity}" based and ``{\it maximum flow}" based.

\noindent {\bf Network connectivity reliability (NCR)}: 
NCR is a classical reliability measure adopted by most work~\cite{jiang2012-reliability_mc,Yeh2010-PSO_MC,Ecuyer2011-ZV_SIS}. The network\!\!\newline
is modelled as a graph where links are either failed~or~operational,
and NCR is measured by the probability that~a~given~set\newline
of nodes are connected when links fail with given probabilities. 
Authors in \cite{jiang2012-reliability_mc} take 
network repair policies into consideration to model link failures, 
and estimate NCR with the classical~MC
Authors in \cite{Yeh2010-PSO_MC} combine MC 
with the particle swarm~optimization to handle the NCR problem. 
To improve the efficiency of MC, authors in \cite{Ecuyer2011-ZV_SIS} apply the IS method and use pre-computed ``\textit{graph minimal cuts}" to approximate the optimal IS estimator.

\noindent {\bf Maximum flow reliability (MFR)}:
Another line of work \cite{botev2014reliability,datta2016reliability,gertsbakh2014permutational}
generalizes the NCR problem by considering link~capac-ities:
link capacities are determined by link statuses,~i.e.,~operational,
failed or partially failed, which follow certain probability distributions.
Given one source and one sink, MFR is defined as
the probability that the maximum flow, 
i.e., the maximum achievable bandwidth from the source to the sink, 
is above a given threshold. 
In~\cite{botev2014reliability}, link capacities are assumed~to be continuous 
and the MC splitting method is applied for~the\newline MFR estimation. 
Authors in \cite{gertsbakh2014permutational} follow the~idea of permutating MC 
and assume all links fail at the beginning and 
each one of them gets repaired after~a random time.
Authors in~\cite{datta2016reliability} consider estimating MFR with
the order minimal cut sets.

\begin{table*}[ht]\footnotesize\centering
\caption{A comparison between classical network reliabilities and flow availability}\label{tb:com}
\resizebox{1.95\columnwidth}{0.215\columnwidth}{
\begin{tabular*}{2.100\columnwidth}{|m{0.200\columnwidth}||m{0.480\columnwidth}|m{0.510\columnwidth}|m{0.707\columnwidth}|}\hline
\SC{0.200\columnwidth}{\bf Reliability\newline Measurement}{|}{||} &\SC{0.480\columnwidth}{\bf NCR~\cite{jiang2012-reliability_mc,Yeh2010-PSO_MC,Ecuyer2011-ZV_SIS,connection_availability,service_availability}}{}{|}
&\SC{0.510\columnwidth}{\bf MFR~\cite{botev2014reliability,gertsbakh2014permutational,datta2016reliability}}{}{|} &\SC{0.707\columnwidth}{\bf Flow Availability}{}{|}\\\hline
\SC{0.200\columnwidth}{\bf Definition}{|}{||} 	
        		& $\mathbb{P}[$node $u$ and $v$ are connected$]$
		& $\mathbb{P}[$max flow from $u$ to $v{\geq}\zeta]$
		& $\mathbb{P}[$flow $f_i$'s demand is satisfied$]$ \\\hline
\SC{0.200\columnwidth}{\bf Required\newline Information}{|}{||}
		& Topology Information: $G\bk{V,\!E}$ and $\pmb{p}$.\!\!
		& Topology Information: $G\bk{V,\!E}$, $\pmb{p}$ and $\pmb{c}$.\!\!\!\!
		& Topology Information: $G\bk{V,\!E}$, $\pmb{p}$ and $\pmb{c}$;\newline
		Flow Information: $F$ and $(\!s_{i},t_{i},d_{i}\!)$ for $\forall f_i{\in}F$.\!\!\!\\\hline
\SC{0.200\columnwidth}{\bf Application\newline scenarios}{|}{||}  	
        & 1) Topology design evaluation.
	& 1) Topology design evaluation;\newline 2) Capacity planning design evaluation.\!\!
	& 1) Topology design evaluation; 2) Capacity planning design evaluation; 3) Sales projection design evaluation.\!\! \\\hline
\SC{0.200\columnwidth}{\bf Relationship}{|}{||}
        & \multicolumn{3}{p{1.785\columnwidth}|}{ {\bf The flow availability is a generalization of network connectivity and maximum flow based reliabilities.}\newline
        When link capacity $c_i{=}\infty$ for $\forall i$, $\mathbb{P}[$flow $f_i$'s demand is satisfied$]{=}\mathbb{P}[$node $u$ and $v$ are connected$]$.\newline
        When flow set $F$ contains only one flow, $\mathbb{P}[$flow $f_i$'s demand is satisfied$]{=}\mathbb{P}[$max flow from $u$ to $v{\geq}\zeta]$.}\\\hline
        \multicolumn{4}{p{2.050\columnwidth}}{{\sl Note}: Consider in network $G(V,E)$, there exist node $u,v$ and flow $f_i$ from $u$ to $v$ with demand $\zeta$. Let vector $\pmb{p}$ and $\pmb{c}$ denote information of failure probability and capacity across all links. Consider in flow set $F$, each flow $f_i$ is associated with a source $s_i$, a destination $t_i$ and a demand $d_i$.} \\
\end{tabular*}}\vspace{3pt}
\end{table*}

\noindent{\bf Other reliabilities}: Some works also study 
the {\it connection availability} \cite{connection_availability} 
and {\it service availability} \cite{service_availability}, 
and consider the probability that a connection or service is available. 
Authors~in\!\newline \cite{connection_availability} 
evaluate the connection availability by 
computing~the~connection probability of a small subset of nodes exactly. 
Authors in \cite{service_availability} evaluate the service availability 
by using IS to estimate path availabilities. 
The problem considered in \cite{connection_availability,service_availability} 
can~be\newline transformed to a problem of determining 
the connectivity of certain nodes, where the network topology is given. Hence,
\cite{connection_availability,service_availability} 
are essentially the same with the NCR related work.
\vspace{-15pt}

%%%%%%%%%%%%%%%%%%%%%%%%%%%%%%%%%%%%%%%%%%
\subsection{\textbf{Comparisons with Classical Reliability Estimation Work}}\label{subsec:related_comp}

We consider the ``flow availability" as our reliability measure.
We first give the definition of the flow demand.
\begin{definition}
The ``{\it demand}" of flow $f$ is the quality of~service\!\!\newline (QoS) 
requirements decided by $f$'s SLA objective.
\end{definition}
Consider different SLA objectives, the flow demand can~be,\newline
for instance, the {\it bandwidth demand}, {\it latency demand} or {\it packet loss demand},
which specifies $f$'s QoS requirement on bandwidth, transmission latency or packet loss.
{\color{black}
To be concrete and so easier to understand,
we take the bandwidth demand as an example,
and the following analysis works the same for other demands.}
We define the flow availability as:
\begin{definition}
For a given {\it topology information}, {\it flow information}, 
{\it routing policy} and {\it resource allocation policy}, 
the ``{\it flow availability}" of $f$ is the probability that $f$'s demand is satisfied.\!\!\!
\end{definition}

Comparing with the state-of-the-art methods, 
our methods have the following advantages:

\noindent 1) The flow availability can be applied to 
{\it evaluate both NCR, given all links have unlimited capacities, and MFR, 
given the network contains only one flow}. 
Yet,~neither~NCR~nor~MFR~can address FAVE. 
To illustrate, consider the example in Fig.\ref{fig:intro_eg}. 
If link 1 fails, the network is still connected but 
neither flow~1~nor 2's demands can be satisfied. 
Also, the maximum flow from~$A$\newline to $B$ still achieves 10 units, 
but it does not imply flow~1~succeeds, 
which depends both on resource~allocation and routing policies and other competing flows.

\noindent 2) Flow availability can be applied to 
{\it evaluate not only the~reli-ability of network designs, 
including topology~design~and~capacity planning, 
but also the feasibility of sales projection}. 
In contrast, NCR only applies to 
the topology design evaluation and 
MFR only applies to the capacity planning~evaluation, 
for they utilize solely the (partial) topology information. 
\onlytech{We demonstrate these applications in our technical report~\cite{tech}.}

To summarize, FAVE generalizes the \onlypaper{NCR and MFR} \onlytech{network reliability} estimations 
and considers a more realistic problem setting.~Moreover, 
it can be applied to evaluate impacts of 
more factors,~e.g.,~network topology, capacity and flow information on network~per-formances, 
and provides more accurate evaluation results.~The 
detailed comparisons can be found in Table. \ref{tb:com}.
\vspace{-3pt}

%%%%%%%%%%%%%%%%%%%%%%%%%%%%%%%%%%%%%%%%%%%%%%%%%%%%%%%%%%%%%%%%%%%
\subsection{\textbf{Sampling Methods for Network Reliability Estimations}}\label{subsec:sampling_basis}\vspace{-1pt}

\noindent {\bf Network reliability estimation problem:} Let the network~be

\noindent modelled as a directional multigraph $G{:=}\bk{V\!{,}E}$
with $N_v$~nodes\newline in node set $V$ and $N_l$ links in {\color{black} link set $E$}.
Each link $e_i{\in}E$~is associated with a tuple $(p_i, c_i, x_i)$
with a small probability~$p_i$~to\newline represent $e_i$'s failure probability, 
a capacity $c_i$, and~a~status~$x_i$,\newline 
where $x_i{=}1$~\bk{$x_i{=}0$} means $e_i$ is down~(up). 
Let $\pmb{p}{=}\{p_1,...,p_{N_l}\!\}$,\!\!\newline 
$\pmb{c}{=}\{c_1,...,c_{N_l}\!\}$ and $\pmb{x}{=}\{x_1,...,x_{N_l}\!\}$ be 
failure probability,~capacity and status across all links respectively. 
$\pmb{x}$, also called the ``{\it failure configuration}", 
is generated by the link failure distribution $p(\pmb{x})$ induced by $\pmb{p}$. 
There are $2^{N_l}$ possible configurations of $\pmb{x}$, 
which is huge for a large realistic network.

Let $\mathcal{R}$ be the indicator function of some interested event~$A$.
According to the reliability definition,
$A$ can refer to the~event\newline
that a subset of nodes are unconnected, or the maximum~flow~is\newline
below the required threshold, 
or, in our case,~the~flow~demand is unsatisfied. 
Given link statuses described by $\pmb{x}$, $\mathcal{R}\bk{\pmb{x}}{=}1$
if $A$ is observed and $\mathcal{R}\bk{\pmb{x}}{=}0$ vice versa. 
The network unreliability, which is the occurrence probability of $A$, 
can~be~computed via the following integral in {\it the discrete measure space}:
\begin{equation}
\label{eq:integral_org}
 \mu{=}\mathbb{E}_p\left[\mathcal{R}\bk{\pmb{x}}\right]
 {=}\textstyle
 \int\mathcal{R}\bk{\pmb{x}} p\bk{\pmb{x}} \mathrm{d}\pmb{x},
\end{equation}
where $\mathbb{E}_p[\cdot]$ means 
taking the expectation over distribution $p\bk{\pmb{x}}$. 
Then, the network reliability can be obtained by $1{-}\mu$.

\noindent {\bf Monte Carlo (MC) simulation:}
The MC simulation draws failure configurations $\pmb{x}$ independently 
from $p\bk{\pmb{x}}$ and estimate $\mu$ with the following MC estimator:
\begin{equation}\label{eq:estimator_mc}
\hat{\mu}_{\emph{MC}}{=}\tfrac{1}{N}
\textstyle\sum_{k{=}1}^{N}\mathcal{R}\bk{\pmb{x}^{\bk{k}}},
\end{equation}
where $N$ is the number of simulation steps 
and $\pmb{x}^{\bk{k}}$ is the~$k$th generated failure configuration.
As MC generates link statuses by true link failure probabilities 
(which can be small),~it~is~rare\newline to observe link failures, 
and even rarer to observe $A$.~This~implies that 
we need a large $N$ to gain the desired accuracy.

\noindent {\bf Importance sampling (IS):}
To improve the efficiency of~MC, IS changes the sampling distribution $p\bk{\pmb{x}}$ to increase the~occurrence of $A$, and assigns each sample~$\pmb{x}$ a weight to recover the unbiasedness. Specifically, it replaces Eq.\bk{\ref{eq:integral_org}} by:
\begin{equation}
\label{eq:integral_is}
\textstyle
 \mu {=}\mathbb{E}_p \!\left[\mathcal{R}\bk{\pmb{x}}\right]
  	{=}{\int}\mathcal{R}\bk{\pmb{x}} \frac{p\bk{\pmb{x}}}{q\bk{\pmb{x}}} q\bk{\pmb{x}} \mathrm{d}\pmb{x}
  	{=} \mathbb{E}_q \!\left[\mathcal{R}\bk{\pmb{x}}\frac{p\bk{\pmb{x}}}{q\bk{\pmb{x}}}\right],
\end{equation}
where $q\bk{\pmb{x}}$ is the ``\textit{importance distribution}". 
For convenience, denote $\omega\bk{\pmb{x}}{=}p\bk{\pmb{x}}/q\bk{\pmb{x}}$ as the weight. 
Therefore, the above~expectation is estimated by the following IS estimator:
\begin{equation}\label{eq:estimator_is}
  \hat{\mu}_{\emph{IS}}
  \textstyle
  {=}\frac{1}{N}\sum_{k{=}1}^{N} \mathcal{R}\bk{\pmb{x}^{\bk{k}}} 
  	\frac{p\bk{\pmb{x}^{\bk{k}}}}{q\bk{\pmb{x}^{\bk{k}}}}
  {=}\frac{1}{N}\sum_{k{=}1}^{N} \mathcal{R}\bk{\pmb{x}^{\bk{k}}}\omega\bk{\pmb{x}^{\bk{k}}}.
\end{equation}

\noindent {\bf Estimator efficiency evaluation:} 
An estimator's efficiency is often measured by its ``{\it variance}". 
Take the MC estimator as~an example, its variance is given by:
\begin{equation}
\label{eq:estimator_mc_var}
   \mathbb{V}_p\left[\hat{\mu}_{\emph{MC}}\right]
   {=}\tfrac{1}{N}\mathbb{V}_p\left[\mathcal{R}\bk{\pmb{x}}\right]
   {=}\tfrac{1}{N}\bk{\mu{-}\mu^2},
\end{equation}
\noindent where $\mathbb{V\!}_p[\cdot]$ means 
taking the variance over the distribution $p\bk{\pmb{x}}$.
The IS estimator's variance can be expressed as:
\begin{equation}
\label{eq:estimator_is_var}
\mathbb{V}_q\left[\hat{\mu}_{\emph{IS}}\right]
{=}\tfrac{1}{N}\mathbb{V}_q\left[\mathcal{R}\bk{\pmb{x}}\omega\bk{\pmb{x}}\right]
{=}\tfrac{1}{N}\left(\mathbb{E}_p\left[\mathcal{R}\bk{\pmb{x}}\omega\bk{\pmb{x}}\right]{-}\mu^2\right).
\end{equation}
Note that the MC estimator is a special case of the IS~estimator
if $p\bk{\pmb{x}}{=}q\bk{\pmb{x}}$, $\forall\pmb{x}$.
Define $\sigma_q^2{=}\mathbb{V}_q\left[\mathcal{R}\bk{\pmb{x}}\omega\bk{\pmb{x}}\right]$ 
as the ``{\it one-run variance}" of the IS estimator. 
To achieve a desired estimation\newline
accuracy, the CI width should be bounded by a threshold~$\delta$,~i.e.,\newline
${2\alpha}{\sigma_q}{/\!}\sqrt{N}{\leq}{\delta}$, 
and the simulation cost is $N{\geq}\bk{2\alpha \sigma_q /\!\delta}^2$,~where
$\alpha$ is a constant decided by the~required~confidence level. 
Thus, {\it a small and bounded $\sigma_q^2$ implies an efficient estimator}.

\begin{table}[htp]\small\centering\vspace{10pt}
\caption{Important Notations}\label{tb:notation}
\resizebox{0.975\columnwidth}{0.360\columnwidth}{
\begin{tabular}{|m{0.16\columnwidth}<{\centering}|m{0.96\columnwidth}|}
\hline
{\bf Notations} &{\bf Descriptions}\\\hline
$N_l, N_f$&The number of transportation links and flows. \\
$e_i$	&The $i$th link, where $i{\in}\{1,{\cdots},N_l\}$.\\
$p_i$, $p_i(\!x_i\!)$, $\pmb{p}$, $p(\pmb{x})$
		&$p_i$ and $p_i(x_i){=}p_i^{x_i}(1{-}p_i\!)^{1\!{-}\!x_i}$ are $e_i$'s failure probability~and~distribution.
		$\pmb{p}{=}(p_1,...,\!p_{N_l}\!)$ and $p(\pmb{x})$ is the distribution induced by~$\pmb{p}$.\!\!\\
$\pmb{x}$	&Failure configuration. $\pmb{x}{=}{(x_1,...,x_{N_l}\!)}$ and $x_i{=}1$ ($x_i{=}0$) means that the link is down~(up).\!\!\\
$f_i$		&The $i$th flow, where $i{\in}\{1,\cdots,N_f\}$.\\
$(s_i,\!t_i,\!d_i,\!o_i)$	&$f_{\!i}$'s flow information, including source $s_i$, destination $t_i$, bandwidth demand $d_i$ and availability target $o_i$.\\
$\mathcal{R}$
  		&Indicator function of traffic engineering simulation. Given $\pmb{x}$, $\mathcal{R}(\pmb{x}){=}1$ ($\mathcal{R}(\pmb{x}){=}0$) if the flow fails~(succeeds).\\
$L,\mathcal{L}$		&$L{\subseteq\!}\left\{1,...,\!N_l\!\right\}$ is a link set. $\mathcal{L}$ is a collection of link sets.\!\!\\
$\Psi(L)$,\; $\Psi^{{-}1}(\pmb{x})$	&$\Psi(L)$ maps a link set $L$ to the failure configuration $\pmb{x}$ which satisfies: $\forall i{\in}L$, $x_i{=}1$; $\forall i{\not\in}L$, $x_i{=}0$. $\Psi^{{-}1}(\pmb{x})$ maps a failure configuration $\pmb{x}$ to the link set $L{=}\{i|x_i{=}1{,}i{\in}\{1,\!{\cdots},\!N_l\}\!\}$.\!\!\\
$\mathrm{span}{(}L{)}$, $\mathrm{span}(\mathcal{L})$
		&$\mathrm{span}(L)$ is the collection of link set $L$'s supersets and $\mathrm{span}(\mathcal{L})$ is the collection of supersets of all link sets $L{\in}\mathcal{L}$.\\
$\mathcal{F}$	&$\mathcal{F}{=}\left\{L | \mathcal{R}(\Psi(L)){=}1, L{\subseteq}\{1,{\cdots},N_l\} \right\}$, the collection of all failure link sets which can result in flow failures.\\
$S,\mathcal{S}$ &A \textit{SEED} $S$ is a special link set satisfying: 1) $S{\in}\mathcal{F}$; 2) $\forall L {\subsetneqq} S$, $L {\not\in} \mathcal{F}$; 3) $\forall L {\supseteq} S$, $L {\in} \mathcal{F}$. And $\mathcal{S}$ is the collection of all SEEDs.\\
  \hline
\end{tabular}}
\end{table}

\noindent {\bf Zero-variance (ZV) importance distribution}:
The following theorem gives an optimal importance distribution $q^*\!\bk{\pmb{x}}$:
\begin{theorem}[Zero-variance importance sampling]\label{thm:is_dist_opt_is}
The IS~estimator in Eq.\bk{\ref{eq:estimator_is}} can achieve zero-variance, 
i.e, $\sigma_q^2{=}0$, 
if~the importance distribution $q\bk{\pmb{x}}{=}q^*\bk{\pmb{x}}$ where:
\begin{equation}\label{eq:is_dist_opt}
  q^*\bk{\pmb{x}}{=}\mathbb{P}\left[\pmb{x}|\mathcal{R}(\pmb{x}){=}1\right].
\end{equation}
\end{theorem}

\noindent {\bf Remark:} 
Although the ZV property implies a minimum simulation cost, 
it is non-trivial and sometimes even impossible to compute 
the ZV importance distribution $q^*\bk{\pmb{x}}$.
Hence, the key of designing an efficient IS estimator lies in 
the approximation of $q^*\bk{\pmb{x}}$. 
As for different applications and problem definitions,\newline 
the auxiliary information we can utilize and 
the way~to~approximate $q^*\bk{\pmb{x}}$ are different, 
designing efficient IS estimators are highly challenging and problem-dependent.

\section{\textbf{Problem Definition}}\label{sec:pro_def_fave}

Consider the network $G\bk{V,E}$ with 
topology information~$\pmb{p},\newline \pmb{c},\pmb{x}$,
as defined in Section \ref{subsec:sampling_basis}. 
We consider a flow set $F$ with $N_f{=}|F|$ flows and each flow $f_i{\in}F$ 
is associated with a tuple $(s_i,t_i,d_i,o_i)$ specifying 
$f_i$'s source $s_i$, destination~$t_i$, demand $d_i$ 
and availability target $o_i$. 
We also define the following:
\begin{definition}
A flow \textbf{fails}~(\textbf{succeeds}) if its demand is unsatisfied (satisfied), 
e.g., the allocated bandwidth cannot (can) support its bandwidth demand.
\end{definition}
\noindent We redefine the function $\mathcal{R}$ to indicate the network routing:
\begin{equation}\label{eq:te_func_02}
\mathcal{R}({\cdot}): (\mathbb{G}, \mathbb{F}, \mathbb{D}, \mathbb{B}) \rightarrow\{0,1\},
\end{equation}
where $\mathbb{G}$ represents the topology information,
including~a~tuple $(p_i,c_i,x_i)$ for every link $e_i{\in}E$; 
$\mathbb{F}$ represents the flow~informa-tion, 
including a tuple $(s_i,t_i,d_i,o_i)$ for every flow $f_i{\in}F$; 
\mk{$\mathbb{D}$ and $\mathbb{B}$ represent 
the underlying routing and resource allocation policies, 
e.g., shortest path policy and max-min fairness policy. 
Function $\mathcal{R}$ outputs~1~if the interested flow fails and 0 other-wise.
We assume~all~information in $\bk{\mathbb{G}, \mathbb{F}, \mathbb{D}, \mathbb{B}}$ is known,~exc-ept link statuses described by $\pmb{x}$. To simplify the expression,~let:}
\begin{equation}\label{eq:te_func}
\mathcal{R}({\cdot}): \pmb{x}\rightarrow\{0,1\}.
\end{equation}
Namely, given all other information, the routing result indicated\!

\noindent by $\mathcal{R}$ only depends on the failure configuration $\pmb{x}$, 
which is generated by the link failure distribution $p\bk{\pmb{x}}$ induced by~$\pmb{p}$.

The unavailability $\mu$, also called the flow~failure probability, 
of a specific flow can be computed via~the integral in Eq.\bk{\ref{eq:integral_org}}.
Our goal is to evaluate availabilities of (all) flows in $F$.~Yet,
the complexity of function $\mathcal{R}$~and the high~dimensionality of the topological space 
make it impossible to evaluate the flow availability $1\!{-}\!\mu$ analytically. 
One alternative is to~estimate~$\mu$~via simulations. 
We name the network~\underline{f}low \underline{av}ailability~\underline{e}stimation problem as ``FAVE". 
And we can show it generalizes~the~classical network reliability estimation problem~\cite{jiang2012-reliability_mc,Yeh2010-PSO_MC,Ecuyer2011-ZV_SIS,Lee2010-crosslayer_reliability,botev2014reliability,gertsbakh2014permutational,datta2016reliability}.
%\vspace{-3pt}
\begin{theorem}\label{thm:fave_gen}
The FAVE problem generalizes both the network connectivity based and 
the maximum flow based network~reliability estimation problems.
\end{theorem}

\noindent{\bf Remark:} 
Due to the page limit,
we only provide the sketch proof of Thm. {\ref{thm:conds_BRE_VRE}} in this paper
and we leave proofs of all other lemmas and theorems in the technical report~\cite{tech}.

\mk{In addition to measuring the estimation efficiency with~the variance,
we also consider two attractive error bound properties:\!\!}
\onlytech{\textit{bounded relative error} and \textit{vanishing relative error} \cite{Ecuyer2011-ZV_SIS}.\!\!}

\begin{definition}[Bounded Relative Error]
\label{def:bre_def}
An estimator with~expectation $\mu$ and variance $\sigma^2/N$ 
has the bounded relative error (BRE) property
{\color{black} 
if $\sigma{=}O\bk{\mu}$, i.e., the coefficient of variation (CV) 
$\varepsilon_{\emph{CV}}{\triangleq}\sigma/\mu$ satisfies
$\lim_{\mu {\rightarrow} 0}\varepsilon_{\emph{CV}}{<}\infty$. }
%is bounded when $\mu {\rightarrow} 0$.
\end{definition}

\begin{definition}[Vanishing Relative Error]
\label{def:bre_def}
An estimator with~expectation $\mu$ and variance $\sigma^2/N$
has the vanishing relative error (VRE) property
{\color{black} 
if $\sigma{=}o\bk{\mu}$, i.e., $\lim_{\mu {\rightarrow} 0}\varepsilon_{\emph{CV}}{=}0$.}
%$\varepsilon_{\emph{CV}}{\triangleq}\sigma/\mu {\rightarrow}0$ when~$\mu{\rightarrow}0$.\!\!
\end{definition}

\noindent {\bf Remark:}
Note that the VRE property is stronger than~the~BRE property.
The variance of MC is $\sigma^2{/}N{=}\bk{\mu{-}\mu^2}{/}N$,
i.e., $\sigma{=} O\bk{\!\sqrt{\mu}}$. This implies that the MC estimator satisfies neither BRE nor VRE property. In the following, we will discuss how to design estimation methods which have above properties.

\section{\textbf{Algorithm Design}}\label{sec:alg_design}
We first describe our design for ``the single flow case". 
We start with a baseline IS design to gain insights for efficient estimations. 
Then we introduce ``SEED" and our~SEED~methods.
\vspace{-10pt}

\subsection{\textbf{A Baseline Importance Sampling Design}}\label{subsec:zv-is}
\noindent{\bf 1) ZV importance distribution approximation}: 
{\color{black} It seems~easy to design an IS estimator}
% At the first glance,
%it seems that the design of the IS estimator is easy 
if we can~well approximate $q^*\bk{\pmb{x}}$ {in Eq.\bk{\ref{eq:is_dist_opt}}}.
Yet, the following discussion shows that this is not an easy task.
We use the KL divergence to measure the similarity between 
$q^*\bk{\pmb{x}}$ {\color{black}and its approximation $q\bk{\pmb{x}}$}.
%our approximation distribution $q(\pmb{x})$. 
We derive $q\bk{\pmb{x}}$ by:
\begin{theorem}\label{thm:is_app}
Assume the optimal importance distribution~$q^*\bk{\pmb{x}}$ 
in Eq.\bk{\ref{eq:is_dist_opt}} is approximated by a product form distribution:
\begin{equation}
	\label{eq:is_dist_prod01}
	q\bk{\pmb{x}}
	{=}\textstyle
	\prod_{i{=}1}^{N_l}q_i\bk{x_i},
\end{equation}
then the KL divergence $\emph{KL}\bk{q^*\|q}$ is minimized when:
\begin{equation}
	\label{eq:kl_dis04}
	q_i\bk{x_i}{=}\mathbb{P}\left[x_i|\mathcal{R}\bk{\pmb{x}}{=}1\right].
\end{equation}
\end{theorem}

\noindent By now, the estimation of 
$\mathbb{P}\left[x_i|\mathcal{R}\bk{\pmb{x}}{=}1\right]$ 
becomes a new~problem. 
In fact, even given the exact $\mathbb{P}\!\left[x_i|\mathcal{R}\bk{\pmb{x}}{=}1\right]$ 
expressions~or\newline values,
the performance of this IS method is still~not~guaran-teed: 
minimizing the KL divergence can only lower~bound~the\newline
estimator's variance, and the lower bound depends on how~well\!\newline
$q\bk{\pmb{x}}$ in Eq.\bk{\ref{eq:kl_dis04}} can approximate $q^*\bk{\pmb{x}}$. 
To see this,~consider~the case where a network 
has 2 nodes connected by~3~parallel~links with $\pmb{p}{=}\bk{0.001, 0.2, 0.001}$. 
The flow fails if~link statuses ${\bf \pmb{x}}$ are 
$\bk{1{,}1{,}0}$, $\bk{1{,}1{,}1}$ or $\bk{0{,}1{,}1}$.
According to Thm. \ref{thm:is_app}, 
one~possible 

\noindent importance distribution is $\pmb{q}{=}\bk{0.50025, 1, 0.50025}$.~In~this~case, 
$q^*\bk{\pmb{x}}$ is {\it not well approximated} by $q(\pmb{x})$ in Table\! \ref{tb:is_appexample}.

\vspace{10pt}
\begin{table}[ht]\small\centering
\caption{An example for the IS method}\label{tb:is_appexample}
\resizebox{0.9\columnwidth}{!}{
\begin{tabular}{|c||lll|}
\hline
$\pmb{x}$ &$(1,1,0)$ &$(1,1,1)$ &$(0,1,1)$\\
\cline{1-4}
$q^*\!\bk{\pmb{x}}{=}\mathbb{P}\left[\pmb{x}|\mathcal{R}\bk{\pmb{x}}{=}1\right]$
  &0.49975   &0.00050   &0.49975\\
$q\bk{\pmb{x}}{=}\prod_{i{=}1}^{N_l}\mathbb{P}\left[x_i|\mathcal{R}\bk{\pmb{x}}{=}1\right]$
  &0.25000   &0.25025   &0.25000\\[1pt]
\hline
\end{tabular}}
\end{table}

{\noindent \bf Remark:}
The above example illustrates that minimizing KL divergence 
cannot provide the IS method a performance guarantee: 
sometimes our chosen $q\bk{\pmb{x}}$ 
can be very different from $q^*\bk{\pmb{x}}$.~A major reason is that
the baseline IS assumes $q\bk{\pmb{x}}$
has the product form in Eq.\bk{\ref{eq:is_dist_prod01}}, 
i.e., it considers link failures are independent 
and ignores the correlations among them. 
We will discuss how to improve the design of $q\bk{\pmb{x}}$ in later sections.

\vspace{5pt}
\noindent {\bf 2) Estimation error bound analysis}: 
The variance bounds of the baseline IS method 
is given by the following theorem:
\begin{theorem}\label{thm:is_variance_bound}
If the IS estimator in Eq.\bk{\ref{eq:estimator_is}} 
takes $q\bk{\pmb{x}}$ in Eq.\bk{\ref{eq:kl_dis04}} 
as its importance distribution, 
then $\!\mathbb{V}_{\!q}{[}\hat{\mu}_{\emph{IS}}{]}\!$ is bounded by:
\begin{equation}\small\label{eq:is_variance_bound}
	\textstyle
	\frac{\mu^2\emph{KL}\bk{q^*\!\|q}}{N}{\leq}
 	\mathbb{V}_q\!\left[\hat{\mu}_{\emph{IS}}\right] {\leq}
 	\frac{\mu^2 \sqrt{2\log2\emph{KL}\bk{q^*\!\|q\,}}}{N \min_{\pmb{x}}q\bk{\pmb{x}}}.
\end{equation}
Here $q^*\!$ is the optimal importance distribution given by Eq.\bk{\ref{eq:is_dist_opt}}.
\end{theorem}

{\noindent \bf Remark:}
Thm. \ref{thm:is_variance_bound} implies that although 
the variance of baseline IS is upper bounded, 
it can be worse than the variance of MC. 
This motivates us to seek for 
more efficient sampling methods with better error bounds.
%Although the naive IS estimator has an upper bounded variance, the upper bound can be worse than the variance of MC. This motivates us to seek for more efficient sampling methods with better error bounds.

\subsection{\textbf{Conditions for Efficient Sampling}}\label{sec:sis-cond}
In the baseline IS design, 
link importance distributions~$q_i\bk{x_i}$\!\newline
are assumed to be independent.
This assumption simplifies~the problem, 
but does not correspond to the reality. 
Consider~the\newline example in Section \ref{subsec:zv-is}, 
$\mathbb{P}[x_3|\mathcal{R}{=}1\!{,}x_1{=}1]$ 
greatly~differs~from\!\newline $\mathbb{P}[x_3|\mathcal{R}{=}1{,}x_1{=}0]$, 
which implies the dependence among~$q_i\bk{x_i}$\newline 
and the correlation of links' statuses cannot~be~ignored.~Next, 
we propose our {``\it sequential importance sampling" (SIS)} based design 
and take this correlation into consideration.

\vspace{3pt}
\noindent{\bf 1) ZV sequential importance sampling}: 
Let us adapt the ZV importance distribution in 
Eq.\bk{\ref{eq:is_dist_opt}} for the SIS method.
Denote ${\bf \pmb{x}}_{1:i}{=}\bk{x_1,x_2,{\cdots},x_i}$ as the status of the first $i$th links.
\begin{theorem}
\label{thm:is_dist_opt_sis}
For the FAVE problem, the IS estimator in~Eq.\bk{\ref{eq:estimator_is}}
achieves the ZV property if 
the importance~distribution~$q\bk{\pmb{x}}{=}$ $q^*\bk{\pmb{x}}$ for the SIS estimator,
where:
\begin{align}
q^*\bk{\pmb{x}}
{=}
{\textstyle\prod}_{i{=}1}^{N_l}q_i\bk{\pmb{x}}
\quad{\rm and}\quad
q_i\bk{\pmb{x}}
{=}
\tfrac{\mathbb{P}\bk{\mathcal{R}{=}1|\pmb{x}_{1:i}}}
{\mathbb{P}\bk{\mathcal{R}{=}1|\pmb{x}_{1:i{-}1}}}p_i\bk{x_i}.
\label{eq:is_dist_opt03}
\end{align}
\end{theorem}\vspace{1pt}

{\noindent \bf Remark:}
Different from baseline IS, 
SIS generates links' statuses in a sequential manner, 
which enables its importance distribution $q_i\bk{\pmb{x}}$ 
to capture the correlation of links' statuses, 
i.e., {\it links' importance distributions are~dependent}.

\vspace{3pt}
\noindent{\bf 2) Conditions for good error bounds:}
To apply the above SIS estimator, 
we need to estimate (or approximate) 
$\mathbb{P}\left[\mathcal{R}{=}1|\pmb{x}_{1:i}\right]$. 
The following theorem states that the above SIS is~robust~even 
if there exists some error when approximating 
$\mathbb{P}[\mathcal{R}{=}1|\pmb{x}_{\!1{:}i}]$.

\begin{theorem}[Conditions for BRE and VRE properties]
\label{thm:conds_BRE_VRE}
The IS estimator in Eq.\bk{\ref{eq:estimator_is}} has the BRE property 
if $\forall i{=}1,{\ldots},N_l$, $\hat{\mathbb{P}}\left[\mathcal{R}{=}1|\pmb{x}_{1:i}\right]$ satisfies:
\begin{equation}\label{eq:bre_req}
\hat{\mathbb{P}}\left[\mathcal{R}{=}1|\pmb{x}_{1:i}\right] 
{=}
O \left(\mathbb{P}\left[\mathcal{R}{=}1|\pmb{x}_{1:i}\right]\right),
\end{equation}
and the VRE property if 
$\forall i{=}1,{\ldots},N_l$, 
$\hat{\mathbb{P}}\left[\mathcal{R}{=}1|\pmb{x}_{1:i}\right]$
satisfies:
\begin{align}\label{eq:vre_req}
\hat{\mathbb{P}}\left[\mathcal{R}{=}1|\pmb{x}_{1:i}\right] 
{=}
\mathbb{P}\left[\mathcal{R}{=}1|\pmb{x}_{1:i}\right]
    {+}o\left(\mathbb{P}\left[\mathcal{R}{=}1|\pmb{x}_{1:i}\right]\right).
\end{align}
\end{theorem}

\noindent{\bf Proof:}
Assume $\mathbb{P}\left[\mathcal{R}{=}1|\pmb{x}_{1:i}\right]{>}0$.
Otherwise there is no need~to generate the corresponding $\pmb{x}$.
Then, Eq.\bk{\ref{eq:bre_req}} is equivalent to:
\begin{align}\label{eq:bre_req02}
\textstyle
\hat{\mathbb{P}}\left[\mathcal{R}{=}1|\pmb{x}_{1:i}\right]
{=}
\mathbb{P}\left[\mathcal{R}{=}1|\pmb{x}_{1:i}\right]\left(\theta_i{+}o\bk{1}\right),
\end{align}
where $\theta_i$ is a constant. Thus, we can derive that:
\begin{align}
\textstyle
\frac{\hat{\mathbb{P}}\left[\mathcal{R}{=}1|\pmb{x}_{1:i}\right]}
 {\hat{\mathbb{P}}\left[\mathcal{R}{=}1|\pmb{x}_{1:i-1}\right]}
{=}\frac{\mathbb{P}\left[\mathcal{R}{=}1|\pmb{x}_{1:i}\right]}
{\mathbb{P}\left[\mathcal{R}{=}1|\pmb{x}_{1:i-1}\right]}
{\cdot}\left(\frac{\theta_i}{\theta_{i-1}}{+}o\bk{1}\right).
\end{align}        
%Therefore,
\begin{align}
q\bk{\pmb{x}}
&{=}\textstyle
{\prod}_{i{=}1}^{N_l}
	\frac{\hat{\mathbb{P}}\left[\mathcal{R}{=}1|\pmb{x}_{1:i}\right]}
	{\hat{\mathbb{P}}\left[\mathcal{R}{=}1|\pmb{x}_{1:i-1}\right]}p\bk{x_i}
	\nonumber\\
&{=}\textstyle
{\prod}_{i{=}1}^{N_l}
	\frac{\mathbb{P}\left[\mathcal{R}{=}1|\pmb{x}_{1:i}\right]}
	{\mathbb{P}\left[\mathcal{R}{=}1|\pmb{x}_{1:i-1}\right]}p\bk{x_i}
	{\cdot}{\prod}_{i{=}1}^{N_l}\left(\frac{\theta_i}{\theta_{i-1}}{+}o\bk{1}\right)
	\nonumber\\
&{=}\textstyle
	q^*\bk{\pmb{x}}{\cdot}\left({\prod}_{i{=}1}^{N_l}\frac{\theta_i}{\theta_{i-1}}{+}o\bk{1}\right).
\end{align}
\begin{align}
\textstyle
\sigma_q^2
{=}
\mu^2\!\int\!\!\frac{q^*\bk{\pmb{x}}^2}{q\bk{\pmb{x}}}\mathrm{d}\pmb{x}{-}\mu^2 
{=}
\mu^2\!\left(\!{\prod}_{i=1}^{N_l}\!\frac{\theta_{i{-}1}}{\theta_i}{-}1{+}o\bk{1}\!\right)\!\!
{=}
O\bk{\mu^2}.
\end{align}
Namely, BRE property is achieved. 
As Eq.\bk{\ref{eq:vre_req}} is a special~case\newline of Eq.\bk{\ref{eq:bre_req02}}
by restricting $\theta_i{=}1$ for all $i$, 
it is similar to show~that\newline Eq.\bk{\ref{eq:vre_req}} implies
$\sigma_q^2{=}o\bk{\mu^2}$, i.e., VRE property is achieved. 
\done

{\noindent \bf Remark}:
Thm. \ref{thm:conds_BRE_VRE} provides important guidelines 
to design~sampling methods with BRE and VRE properties: 
the estimation $\hat{\mathbb{P}}\left[\mathcal{R}{=}1|\pmb{x}_{1:i}\right]$
should satisfy the conditions listed in Thm. \ref{thm:conds_BRE_VRE}.

\subsection{\textbf{SEED Algorithms}}
\noindent{\bf 1) SEED set and related definition}:
Before introducing how~to\!\newline approximate 
$
	\mathbb{P}\left[\mathcal{R}{=}1|\pmb{x}_{1{:}i}\right]
$
for the SIS estimator,
we first~present\!\newline some definitions used in the later discussion.

Let 
$
\Omega{\triangleq}\left\{\!1,...,N_l\!\right\}
$
be the index set of all links.
We use~function $\Psi\bk{\cdot}$ and $\Psi^{{-}1}\bk{\cdot}$ 
for transformations between the failure configuration $\pmb{x}$ 
and the set of failure links $L{\subseteq}\Omega$:
\begin{align}\label{eq:psi_func}
&\Psi\bk{\cdot}: L {\rightarrow} \pmb{x}, \;
	\mathrm{where}\; 
	\forall i {\in} L,\; x_i{=}1;\;
	\forall i {\not\in} L,\;  x_i{=}0. 
\\[-3pt]
&\Psi^{-1}\bk{\cdot}: \pmb{x} {\rightarrow} L, \; 
	\mathrm{where}\; 
	L {=}\{i|x_i{=}1\}.
\end{align}
For one interested flow, 
denote the collection of all link sets 
whose failures can result in the flow failure by:
\begin{equation}\label{eq:fail_conf}
\mathcal{F}
{\triangleq}
\left\{L |\; \mathcal{R}(\Psi(L)){=}1, L{\subseteq}\Omega\right\}.
\end{equation}
Also, denote the collection of all supersets of link set $L$ by:
\begin{equation}
\mathrm{span}\left(L\right)
{\triangleq}
\left\{L^\prime|L{\subseteq}L^\prime{\subseteq}\Omega\right\}.
\end{equation}
Accordingly, for a collection of link sets 
$\mathcal{L}{=}\left\{L\right\}$,
\begin{equation}
\mathrm{span}\left(\mathcal{L}\right)
{\triangleq}
\textstyle
{\bigcup}_{L{\in}\mathcal{L}}\mathrm{span}\left(L\right).
\end{equation}
The probability of the event that ``all links in $L$ fail" is:
\begin{equation}\label{prob_func}
\textstyle
\Phi\bk{L}
{\triangleq}
	{\prod}_{i{\in}L}p_i
{=}
	{\sum}_{L^\prime{\in}\mathrm{span}(L)}\mathbb{P}\left[\Psi\left(L^\prime\right)\right].
\end{equation}

\begin{definition}[SEED]
\label{def:seed}
Define the \textit{SEED} $S$ as a special link~set 
which satisfies the following conditions:
$$
{\rm (1)} S{\in}\mathcal{F};\qquad
{\rm (2)} {\forall} L{\subsetneqq}S, L {\not\in} \mathcal{F};\qquad
{\rm (3)} {\forall} L {\supseteq}S, L {\in}\mathcal{F}.
$$
We also denote the collection of all SEEDs by $\mathcal{S}$.
\label{def:seed}
\end{definition}
\begin{definition}[Conditional SEED]
Consider the statuses of some links are specified by $\pmb{x}_{1{:}i}$, 
define the \textit{conditional SEED} (\textit{cond-SEED} for short) $S(\pmb{x}_{1{:}i})$ 
as a special link set satisfying:

\vspace{-1pt}
\noindent\resizebox{1\columnwidth}{!}{\begin{minipage}{\columnwidth}\begin{align*}
&\bk{1}
	S\bk{\pmb{x}_{1:i}}{\subseteq} \{i{+}1,{\cdots}, N_l\};   			
&\bk{3}
	{\forall} L{\subsetneqq}S\left(\pmb{x}_{1{:}i}\right), 
	L{\cup}\Psi^{{-}1}(\pmb{x}_{1{:}i}){\not\in}\mathcal{F};
\\[-4.5pt]
&\bk{2}
	S\bk{\pmb{x}_{1:i}}{\cup}\Psi^{{-}1}\bk{\pmb{x}_{1{:}i}}{\in}\mathcal{F};
&\bk{4} 
	{\forall} L{\supseteq}S\left(\pmb{x}_{1{:}i}\right), 
	L{\cup}\Psi^{{-}1}(\pmb{x}_{1{:}i}) {\in}\mathcal{F}.
\end{align*}\end{minipage}}
\vspace{-1pt}

\noindent Also, we denote the collection of all cond-SEEDs 
by $\mathcal{S}\left(\pmb{x}_{1:i}\right)$.
\label{def:seed}
\end{definition}
\vspace{2pt}

{\noindent \bf Examples}: 
We give some examples for the above definitions. 
Consider the example in Section \ref{sec:intro}, 
we have $\mathcal{S}{=}\{\{1\}{,}\{4{,}5\}\!\}$.
By the definition of SEED, 
the failure of any subset of a SEED, 
e.g., $L{=}\{4\}$ or $L{=}\{5\}$, 
will not result in flow 1's failure; 
and the failure of any superset of a SEED, 
e.g., $L{=}\{1{,}2\}$, will result in flow 1's failure.
If given the first four links' statuses by $\pmb{x}_{1{:}4}{=}\bk{0,1,0,1}$, 
there is only one cond-SEED $S\bk{\pmb{x}_{1{:}4}}{=}\{5\}$.

Next, we use ``SEED" and ``SEED set" to capture 
the~correlation of link failures and approximate 
$\mathbb{P}\!\left[\mathcal{R}{=}1|\pmb{x}_{1:i}\right]$ in~Thm.\!~\ref{thm:is_dist_opt_sis}.

\noindent{\bf 2) SEED based IS algorithms}:
Thm.\!~\ref{thm:is_dist_opt_sis} presents 
the optimal importance distribution $q^*\!\bk{\pmb{x}}$ for the SIS method. 
We next~show\!\newline that $q^*\bk{\pmb{x}}$ can be computed exactly 
via Algorithm \ref{alg:seed-zv}.

\begin{algorithm}[!t]
\caption{\small{SEED Based ZV Sampling~(SEED-ZV)}}
\label{alg:seed-zv}
\footnotesize
\begin{algorithmic}[1]
\Require The collection of all SEEDs $\mathcal{S}$
\Ensure An importance distribution $q\bk{\pmb{x}}$ 
	to achieve the ZV property and a sample of failure configuration $\pmb{x}$.
\For{$i {=} 1$ to $N_l$}
    \State $x_i \leftarrow 1$ and $\hat{\mathbb{P}}\left[\mathcal{R}{=}1|\pmb{x}_{1:i}\right] {\leftarrow} 0$;
    \For{all $\emptyset {\neq} A {\subseteq} \mathcal{S}\bk{\pmb{x}_{1:i}} $}
       \State 
       \resizebox{.85\columnwidth}{!}{
	\begin{minipage}{\columnwidth}
	\begin{align*}
        \hat{\mathbb{P}}\left[\mathcal{R}{=}1|\pmb{x}_{1:i}\right] 
        {\leftarrow}
        \hat{\mathbb{P}}\left[\mathcal{R}{=}1|\pmb{x}_{1:i}\right]
        {+}
        \bk{{-}1}^{|A|{-}1} \Phi\left({\bigcup}_{S\bk{\pmb{x}_{1:i}}{\in}A} S\bk{\pmb{x}_{1:i}}\!\right);
        \end{align*}
        \end{minipage}
        }
    \EndFor
    \State Keep $x_i{=}1$ with 
    		$
		q_i\bk{\pmb{x}}
		{=}
		\frac{\hat{\mathbb{P}}\bk{\mathcal{R}{=}1|\pmb{x}_{1:i}}}
			{\hat{\mathbb{P}}\bk{\mathcal{R}{=}1|\pmb{x}_{1:i{-}1}}}p_i
		$;
    \State $
    		\mathcal{S}\bk{\pmb{x}_{1:i}}\gets
		$
		\Call{UpdateCondSEED}{$i,x_i,\mathcal{S}\bk{\pmb{x}_{1:i{-}1}}$}
\EndFor
\Function{UpdateCondSEED}{$i,x_i,\mathcal{S}\bk{\pmb{x}_{1:i{-}1}}$}
   \If{$x_i=1$}
        \State \Return{$\mathcal{S}\bk{\pmb{x}_{1:i}}
        		\gets\left\{ L {\setminus} \{i\}|\; L{\in}\mathcal{S}\bk{\pmb{x}_{1:i{-}1}} \right\}$}
    \Else
        \State \Return{$\mathcal{S}\bk{\pmb{x}_{1:i}}
        		\gets\left\{ L |\; L{\in}\mathcal{S}\bk{\pmb{x}_{1:i-1}}, i {\not\in} L \right\}$}
    \EndIf
\EndFunction
\end{algorithmic}
\end{algorithm}

\begin{theorem}\label{thm:seed_zv_alg}
If failure configurations $\pmb{x}$ are generated using SEED-ZV, 
the estimator in Eq.\bk{\ref{eq:estimator_is}} has the ZV property.
\end{theorem}

{\noindent \bf Remark:} 
Though SEED-ZV has the ZV property, 
it~is~compu-tationally expensive: 
as the need for traversing all combinations of cond-SEEDs 
$S\bk{\pmb{x}_{1{:}i}}{\in}\mathcal{S}$ for each link $e_i$,
the computational complexity is $O\bk{N_l{\cdot}2^{\left|\mathcal{S}\right|}}$.

Note that there is a tradeoff between estimation accuracy~and 
computational complexity to estimate $q^*\!\bk{\pmb{x}}$. 
Next, we~consider sacrificing some estimation accuracy of $q^*\!\bk{\pmb{x}}$, 
i.e., {\it utilize probabilities of the most important cond-SEEDs} 
for estimations (in line 3 of Algorithm \ref{alg:seed-bre}), 
and propose SEED-BRE, 
which has a lower linear computational complexity. 
We show SEED-BRE has the BRE property by the following theorem.
\begin{theorem}\label{thm:seed_bre_alg}
If failure configurations $\pmb{x}$ are generated using SEED-BRE, 
the estimator in Eq.\bk{\ref{eq:estimator_is}} has the BRE property.
\end{theorem}

{\noindent \bf Remark:} 
The computational complexity of SEED-BRE is 
only $O\bk{N_l\!\left|\mathcal{S}\right|}$, 
as it needs to traverse all cond-SEEDs 
in $\mathcal{S}\bk{\pmb{x}_{1:i}}$,~for each $e_i$. 
The size of $\mathcal{S}\bk{\pmb{x}_{1{:}i\!}}$ decreases when $i$ increases.

To further improve estimation accuracy, 
we {\it utilize the~probability sum of cond-SEEDs for estimations} 
(in line 3 of~Algo-rithm \ref{alg:seed-vre}), 
and propose~the SEED-VRE algorithm. 
We next show that SEED-VRE has the VRE property 
if link failures are rare.

\begin{theorem}
\label{thm:seed_vre_alg}
When link failure probabilities are small and 
in the form of $p_i{=}O\bk{\epsilon}, \forall i{\in}\Omega$. 
If failure configurations $\pmb{x}$ are generated by SEED-VRE, 
the estimator in Eq.\bk{\ref{eq:estimator_is}} has the VRE property.
\end{theorem}

{\noindent \bf Remark:} 
The computational complexity of SEED-VRE is also 
$O\bk{N_l\! \left|\mathcal{S}\right|}$, 
as it needs to traverse all the cond-SEED sets in 
$\mathcal{S}\bk{\pmb{x}_{1:i}}$, for each $e_i$.
%\begin{center}
%\begin{minipage}{.90\linewidth}
\begin{algorithm}[t]
\caption{\small{SEED Based BRE Sampling~(SEED-BRE)}}
\label{alg:seed-bre}
\footnotesize
\begin{algorithmic}[1]
\Require The collection of all SEEDs $\mathcal{S}$
\Ensure An importance distribution $q\bk{\pmb{x}}$ to achieve 
		the BRE property and a sample of failure configuration $\pmb{x}$.
\For{$i {=} 1$ to $N_l$}
    \For{$x_i$ in $\{0,1\}$}
        \State 
        	$
        	\hat{\mathbb{P}}\left[\mathcal{R}{=}1|\pmb{x}_{1:i}\right]
        	{\leftarrow}
        	{\max}_{S\bk{\pmb{x}_{1:i}}{\in}\mathcal{S}\bk{\pmb{x}_{1:i}}} 
        	\Phi\bk{S\bk{\pmb{x}_{1:i}}}
        	$;
        \State 
        	$
		q_i\bk{x_i} 
        	{\leftarrow}
		\frac{\hat{\mathbb{P}}\bk{\mathcal{R}{=}1|\pmb{x}_{1:i}}} 
			{\hat{\mathbb{P}}\bk{\mathcal{R}{=}1|\pmb{x}_{1:i-1}}}p_i\bk{x_i}$;
    \EndFor
    \State Normalize $q_i\bk{\pmb{x}}$ by 
    		$q_i\bk{1}{\leftarrow}\frac{q_i\bk{1}}{q_i\bk{1}{+}q_i\bk{0}}$, 
		$q_i\bk{0}{\leftarrow} 1{-}q_i\bk{1}$;
    \State Set $x_{i}$ as 1 with probability $q_i\bk{1}$ and 0 with probability $q_i\bk{0}$;
    \State $\mathcal{S}\bk{\pmb{x}_{1:i}}\gets$
    		\Call{UpdateCondSEED}{$i,x_i,\mathcal{S}\bk{\pmb{x}_{1:i{-}1}}$}
\EndFor
\end{algorithmic}
\end{algorithm}

\begin{algorithm}[!t]
\caption{\small{SEED Based VRE Sampling~(SEED-VRE)}}
\label{alg:seed-vre}
\footnotesize
\begin{algorithmic}[1]
\Require The collection of all SEEDs $\mathcal{S}$
\Ensure An importance distribution $q\bk{\pmb{x}}$ to achieve 
		the VRE property and a sample of failure configuration $\pmb{x}$.
\For{$i {=} 1$ to $N_l$}
    \For{$x_i$ in $\{0,1\}$}
        \State 
        $
        	\hat{\mathbb{P}}\left[\mathcal{R}{=}1|\pmb{x}_{1:i}\right]
		{\leftarrow}
        	{\sum}_{S\bk{\pmb{x}_{1:i}}{\in}\mathcal{S}\bk{\pmb{x}_{1:i}}} 
		\Phi\bk{S\bk{\pmb{x}_{1:i}}}
	$;
        \State 
        $
        	q_i\bk{x_i}
		{\leftarrow}
		\frac{\hat{\mathbb{P}}\bk{\mathcal{R}{=}1|\pmb{x}_{1:i}}}
			{\hat{\mathbb{P}}\bk{\mathcal{R}{=}1|\pmb{x}_{1:i{-}1}}}p_i\bk{x_i}
	$;
    \EndFor
    \State Normalize $q_i\bk{\pmb{x}}$ by 
    $
    		q_i\bk{1} 
		{\leftarrow} 
		\frac{q_i\bk{1}}{q_i\bk{1}{+}q_i\bk{0}}
    $, $q_i\bk{0} {\leftarrow} 1{-}q_i\bk{1}$;
    \State Set $x_{i}$ as 1 with probability $q_i\bk{1}$ and 0 with probability $q_i\bk{0}$;
    \State $\mathcal{S}\bk{\pmb{x}_{1:i}}
    		\gets$\Call{UpdateCondSEED}{$i,x_i,\mathcal{S}\bk{\pmb{x}_{1:i{-}1}}$}
\EndFor
\end{algorithmic}
\end{algorithm}

By now, we have three SEED algorithms, 
i.e., SEED-ZV,\newline SEED-BRE and SEED-VRE, 
to compute the importance distributions of SIS estimator. 
They can achieve ZV, BRE~and~VRE properties respectively, 
and with the computational complexities 
$O\bk{N_l{\cdot} 2^{\left|\mathcal{S}\right|}}$, 
$O\bk{N_l\left|\mathcal{S}\right|}$ 
and $O\bk{N_l\left|\mathcal{S}\right|}$ respectively. 
However, all above discussions focus on the~``single flow~case". 
Next, we will generalize our methods to handle multiple flows 
and take other practical issues into~consideration.\!\!

\section{\textbf{Practical Consideration}}\label{sec:sis-prac}
Previous discussions illustrate that SEED methods work~well\!\newline
in estimating a single flow's availability with 
the full~informa-\newline tion of $\mathcal{S}$. 
Next, we consider more practical issues.
First,~as there are $O\bk{{N_v}^{\!\!2}}$ flows in the network, 
it~is~costly to design a ``customized" estimator for each flow~and 
individually estimate\newline their availabilities. 
If the designed estimator works for a group of flows, 
the computational cost can be reduced significantly. 
Furthermore, as SEED methods rely on the SEED set~which may be 
difficult to obtain the full information at times, 
we consider the case that 
only a partial information of $\mathcal{S}$ is available, 
e.g., we only know some frequently observed SEEDs.\vspace{-3pt}

\subsection{\textbf{Generalization to Multiple Flows Case}}
To provide efficient and accurate availability estimations~for 
a group of flows at the same time, 
one possibility is to~utilize SEED methods to 
design a \textit{pure importance distribution}~$q^{\bk{k}}\bk{\pmb{x}}$\!\newline 
for each flow $f_k{\in}F$, 
then take a mixture of these~pure~distributions 
with a strategy $\mathcal{M}$ to simulate link failures:
\begin{equation}\label{eq:mix_sis_dist01}
	q\bk{\pmb{x}}{=}\mathcal{M}\bk{q^{\bk{1}}\bk{\pmb{x}},\ldots,q^{\bk{N_f}}\bk{\pmb{x}}}.
\end{equation}
To derive such a \textit{mixture importance distribution}, 
we take a weighted sum of these pure distributions:
\begin{equation}\label{eq:mix_sis_dist02}
q\bk{\pmb{x}} {=} {\textstyle\sum}_{k} w_k q^{\bk{k}}\bk{\pmb{x}}, \textstyle\sum_{k} w_k{=}1.
\end{equation}
Here $w_k$ can be viewed as 
the probability of taking $q^{\bk{k}}\!\bk{\pmb{x}}$~to\newline generate $\pmb{x}$.
Denote $\mu^{\bk{k}}$ as $f_k$'s failure probability 
and~$\bk{\sigma_{q}^{\bk{k}}}^2$\newline as 
the one-run variance when taking $q\bk{\pmb{x}}$ 
to estimate $f_k$'s~failure probability. 
Next we analyze error bounds of this mixture\newline sampling strategy, 
when applying to the multiple flows~case.
\begin{theorem}\label{thm:mix_sis_bre}
Using the mixture sampling strategy in Eq.\bk{\ref{eq:mix_sis_dist02}} 
with pure distributions $q^{\bk{k}}\bk{\pmb{x}}$ generated by SEED methods, 
the IS estimator achieves the BRE property for all flows availability estimations.
Specifically, for flow $f_k$, the estimator's one-run variance satisfies:
\begin{equation}\label{eq:mix_sis_var}
\textstyle
\left(\sigma_{q}^{\bk{k}}\right)^2
{\leq}
\left(\tfrac{1}{w_k}{-}1\right)\big(\mu^{\bk{k}}\big)^2
{+}
\tfrac{1}{w_k}\Big(\sigma_{q^{\bk{k}}}^{\bk{k}}\Big)^2.
\end{equation}
\end{theorem}

{\noindent \bf Remark:} 
Thm. \ref{thm:mix_sis_bre} states that, 
when extending to the~multiple flows case, 
our methods guarantee the estimation efficiency~for all flows. 
Designing proper or even optimal weights $\!\{\!w_k\!\}$ is challenging. 
Online learning is a good approach to find a more efficient weight setting, 
and we leave this as a future work.

\subsection{\textbf{Partial Seed Set Information}}
At times, it may be difficult to obtain a ``full" SEED set~$\mathcal{S}$, 
especially when the network is large and flow failures are~rare. 
To provide robust estimations, 
consider the case that 
we have only a partial information of $\mathcal{S}$, 
e.g., limited historical data~of\newline flow failures 
which gives $\mathcal{S}^\prime {\subset} \mathcal{S}$. 
Denote the cond-SEED~set induced by $\mathcal{S}^\prime$ 
and $\pmb{x}_{1:i}$ as $\mathcal{S}^\prime\bk{\pmb{x}_{1{:}i}}$. 
To analyze error bound properties of SEED algorithms, 
we provide the following~lemma.

\begin{lemman}
\label{lm:part_seed_alg}
Given a partial SEED set $\mathcal{S}^\prime\bk{\pmb{x}_{\!1{:}i}}$,
when estimating\newline 
$
	\mathbb{P}\!\left[
		\Psi^{{-}1}\bk{\pmb{x}}
		{\in}
		\mathrm{span}\!\left(\mathcal{S}^\prime \bk{\pmb{x}_{1{:}i}}\right)\!|\pmb{x}_{1{:}i}
	\right]
$:
SEED-ZV and SEED-BRE have ZV and BRE properties respectively;
assume link failure probabilities are small and 
follow the form of $p_i{=}O\bk{\epsilon}, \forall i{\in} \Omega$, 
SEED-VRE have the VRE property.
\end{lemman}

\noindent{\bf Remark}:
Thm.\!~\ref{thm:is_dist_opt_sis} states that 
the estimation~accuracy~depends~on\newline
how well $\mathbb{P}\left[\mathcal{R}{=}1|\pmb{x}_{1:i}\right]$ is approximated.
Given $\mathcal{S}$, 
SEED methods have good error bound properties 
for they can well approximate 
$
\mathbb{P}\!\left[
	\Psi^{\!{-}1}\bk{\pmb{x}}
	{\in}
	\mathrm{span}\bk{\mathcal{S}\bk{\pmb{x}_{1{:}i}}}|\pmb{x}_{1{:}i}
\right]\!
{=}
\mathbb{P}\left[
	\mathcal{R}{=}1
	|\pmb{x}_{1:i}
\right]
$.~However,\newline given a partial SEED set $\mathcal{S}^\prime$, 
the bias between 
$\mathbb{P}\left[\mathcal{R}{=}1|\pmb{x}_{1:i}\right]$ 
and
$
\mathbb{P}\left[
	\Psi^{{-}1}\bk{\pmb{x}}
	{\in}
	\mathrm{span}\!\left(\mathcal{S}^\prime \bk{\pmb{x}_{1{:}i}}\right)\!|\pmb{x}_{1{:}i}
\right]
$ 
should be considered. 
Let us consider the following two cases:
\begin{enumerate}[label=\arabic*),leftmargin=3mm]
\item {\bf Good coverage case}.
SEED methods maintain good error bound properties 
if $\mathcal{S}^\prime$ has a \textit{good coverage}, 
which is defined formally as:
\begin{equation}\small
\mathbb{P}\!\left[
	\Psi^{{-}\!1}\bk{\pmb{x}}
	{\in}
	\mathrm{span}\bk{\mathcal{S}^\prime}
\right]\!
{=}
\mathbb{P}\!\left[
	\Psi^{{-}1}\bk{\pmb{x}}
	{\in}
	\mathrm{span}(\mathcal{S})
\right]\!(1{+}o(1)).
\end{equation}
Here is an example of such a partial SEED set:
\begin{equation}
\label{eq:eg_seed_set}
\!\mathcal{S}^\prime
{=}
\!\left\{S| S{\in}\mathcal{S}, \left|S\right|{\leq}{\min}_{S_j\in\mathcal{S}}|S_j|{+} k\right\}\!, 
{\rm where}\; p_i{=}O\bk{\epsilon}, k{\geq}0.\nonumber
\end{equation}
\item {\bf Poor coverage case}.
$\mathcal{S}^\prime$ may not have a good coverage,~e.g.,\newline 
without prior knowledge of network and flow failures, 
$\mathcal{S}^\prime{=}\emptyset$. 
In such case, the SEED set information can be collected 
via pre-samplings and updated while simulating flow failures.
\onlytech{We provide one possible way to collect SEED set~information in the technical~report \cite{tech}.}
\end{enumerate}

%%%%%%%%%%%%%%%%%%%%%%%%%%%%%%%%%%%%%%%%%%%%%%%%%%%%%%%%%%%%%%%%%%%%%%%%%%%%%%%%%%%%%%%%%%%%%%%%%%%%%%%%%%%%%%%%%%%%%%%%%%%%%
\onlytech{
\subsection{\textbf{SEED Set Collection}}

\begin{algorithm}[t]
\caption{\small{SEED Sets Updating Algorithm (SEED-Updating)}}
\label{alg:seed-updating}
\footnotesize
\begin{algorithmic}[1]
\Require $\{\mathcal{L}^{(k)}\}_{k{=}1}^{N_f}$, where $\mathcal{L}^{(k)}{\subset}\mathcal{F}^{(k)}$ and $\mathcal{L}$ is a ``sperner family"; $N$, the number of simulations.
\Ensure $\{\mathcal{S}^{\prime (k)}\}_{k=1}^{N_f}$.
\For{$i$ = 1 to $N$}
    \State Generate topology sample $\pmb{x}$ according to $q(\pmb{x})$;
    \For{ $k$ = 1 to $N_f$}
        \If{$\mathcal{R}^{(k)}(\pmb{x}){=}1$ and $\Psi^{-1}(\pmb{x}){\not\in} \mathrm{span}\left(\mathcal{L}^{(k)}\right)$}
            \State $\mathcal{L}^{(k)}\leftarrow\mathcal{L}^{(k)}\cup\{\Psi^{-1}(\pmb{x})\}\setminus \{L|L{\in}\mathcal{L}^{(k)},L{\supset}\Psi^{-1}(\pmb{x})\}$;
        \EndIf
    \EndFor
\EndFor
\State $\mathcal{S}^{\prime(k)}\leftarrow \mathcal{L}^{(k)}$ where $1{\leq}k{\leq}N_f$.
\end{algorithmic}
\end{algorithm}

The SEED sets information can be collected via~pre-samplings or updated adaptive while simulating flow failures. In either case, SEED sets need to be updated according~to~the observed simulation results. We provide Alg.~\ref{alg:seed-updating} to conduct this update and claim that Alg.\ref{alg:seed-updating} works via the following theorem:
\begin{theorem}
\label{thm:seed_update}
The SEED-Updating algorithm guarantees that ${\rm span}(\mathcal{L}^{(k)}) \uparrow {\rm span}(\mathcal{S}^{(k)}$, i.e., ${\rm span}(\mathcal{L}^{(k)})$ monotone converges to ${\rm span}(\mathcal{S}^{(k)}$, for $\forall {f_k}{\in}{F}$.
\end{theorem}

\noindent{\bf Remark:} Thm.\ref{thm:seed_update} implies that SEED-Updating can: (1) Improve the coverage quality of the partial SEED set. (2) Guarantee the updated partial SEED set converges to the full SEED set. As the sampling distribution $q(\pmb{x})$ is not specified in Alg.\ref{alg:seed-updating}, SEED-Updating can cooperate with any sampling methods to collect SEED set information. However, the convergence rate depends on the way we generate $\pmb{x}$. We leave the efficient SEED set collection as the future work.}

\section{\textbf{Evaluation of SEED methods}}
\label{sec:simulation}
We evaluate our methods on 
both an illustrative small~scale network and a realistic network 
with topology and traffic~matrices 
extracted from the Abilene network~\cite{Abilene}. 
The simulation cost to guarantee 
the estimation error (or relative error) below~a constant $\delta$ is 
$N{\geq}\frac{4\alpha^2}{\delta^2}\sigma_q^2$ 
(or $N{\geq}\frac{4\alpha^2}{\delta^2}(\!\frac{\sigma_q}{\mu}\!)^2$). 
Hence, we use~{\it one-run variance} $\sigma_q^2$ 
and {\it coefficients of variation} (CV) 
$\textstyle\varepsilon_{\emph{CV}}{=}\frac{\sigma_q}{\mu}$ 
to quantify the estimation efficiency. 
The variance reduction, i.e., $\textstyle\frac{\sigma_{\rm MC}^2}{\sigma_{\rm SEED}^2}$, 
can imply the simulation cost reduction, 
i.e.,~$\frac{N_{\rm MC}}{N_{\rm SEED}}$.

\subsection{\textbf{Experiments on an Illustrative Network}}
The illustrative network\footnote{This is provided so that readers can simulate and validate our methods.} demonstrates the ``best achievable theoretical improvements" using our method, 
compared with MC and baseline IS. 
We~start with the single flow~case, 
where full SEED sets or partial SEED sets 
with good coverage are provided. 
Then we extend~it to the multiple flows case.

\begin{table*}[!tbp]\centering
\caption{Flow unavailability analysis results for the network in Fig.\ref{fig:toy-case_topo}}\label{tb:toy-case}\small\vspace{-1pt}
%\resizebox{1.8\columnwidth}{!}{
\resizebox{1.8\columnwidth}{0.240\columnwidth}{
\begin{tabular*}{1.975\columnwidth}{|p{1.0cm}|p{0.45cm}||p{1.4cm}p{1.4cm}p{1.4cm}p{1.4cm}p{1.6cm}|p{1.2cm}p{1.4cm}p{1.4cm}|}
\hline
\multicolumn{2}{|l||}{\multirow{2}{*}{}}		& \multicolumn{5}{c|}{{\bf Full SEED Set}}		& \multicolumn{3}{c|}{\bf Partial SEED Set}\\
\cline{3-10}
\multicolumn{2}{|l||}{}		& \multicolumn{1}{c}{MC}		& \multicolumn{1}{c}{IS}		
					& \multicolumn{1}{c}{SEED-ZV}		& \multicolumn{1}{c}{SEED-BRE}		& \multicolumn{1}{c|}{SEED-VRE}		
					& \multicolumn{1}{c}{SEED-ZV}		& \multicolumn{1}{c}{SEED-BRE}		& \multicolumn{1}{c|}{SEED-VRE} \\ \hline
\multirow{3}{*}{$\epsilon{=}0.05$}&$\!\!\mu$	&$\!\!3.808{\times}10^{{-}3}$		& $3.808{\times}10^{{-}3}$		
					& $3.808{\times}10^{{-}3}$		& $3.808{\times}10^{{-}3}$			& $3.808{\times}10^{{-}3}$
					& $\!\!3.520{\times}10^{{-}3}$    		& $3.520{\times}10^{{-}3}$     			& $3.520{\times}10^{{-}3}$ \\
                                		&$\!\!\sigma_q^2$  	&$\!\!3.793{\times}10^{{-}3}$ 		& $7.549{\times}10^{{-}5}$
					& $0$                        & $1.663{\times}10^{{-}6}$     		& $1.059{\times}10^{{-}8}$
					& $\!\!0$                    & $1.145{\times}10^{{-}7}$     		& $1.277{\times}10^{{-}9}$\\
                          		&$\!\!\sigma_q{/}\mu$&$\!\!1.617{\times}10^{1}$    		& $2.281$
					& $0$                         & $3.387{\times}10^{{-}1}$     		& $2.702{\times}10^{{-}2}$
					& $\!\!0$                     & $8.886{\times}10^{{-}2}$     		& $9.384{\times}10^{{-}3}$\\
\hhline{|=|=||=====|===|}
\multirow{3}{*}{$\epsilon{=}0.01$}&$\!\!\mu$   	&$\!\!1.419{\times}10^{{-}4}$ 		& $1.419{\times}10^{{-}4}$
					& $1.419{\times}10^{{-}4}$    		& $1.419{\times}10^{{-}4}$     			& $1.419{\times}10^{{-}4}$
					& $\!\!1.393{\times}10^{{-}4}$    		& $1.393{\times}10^{{-}4}$    	 		& $1.393{\times}10^{{-}4}$ \\
                                		&$\!\!\sigma_q^2$  	&$\!\!1.418{\times}10^{{-}4}$ 		& $1.323{\times}10^{{-}7}$
					& $0$                        & $6.586{\times}10^{{-}10}$    		& $1.523{\times}10^{{-}13}$
					& $\!\!0$                    & $4.282{\times}10^{{-}11}$   		& $1.743{\times}10^{{-}14}$\\
                          		&$\!\!\sigma_q{/}\mu$&$\!\!8.395{\times}10^{1}$    		& $2.564$
					& $0$                         & $1.809{\times}10^{{-}1}$     		& $2.751{\times}10^{{-}3}$
					& $\!\!0$                     & $4.697{\times}10^{{-}2}$     		& $9.477{\times}10^{{-}4}$\\
\hhline{|=|=||=====|===|}
\multirow{3}{*}{$\epsilon{=}0.001$}&$\!\!\mu$  	&$\!\!1.392{\times}10^{{-}6}$ 		& $1.392{\times}10^{{-}6}$
					& $1.392{\times}10^{{-}6}$    		& $1.392{\times}10^{{-}6}$     			& $1.392{\times}10^{{-}6}$
					& $\!\!1.389{\times}10^{{-}6}$    		& $1.389{\times}10^{{-}6}$     			& $1.389{\times}10^{{-}6}$ \\
                                		&$\!\!\sigma_q^2$  	&$\!\!1.392{\times}10^{{-}6}$ 		& $1.348{\times}10^{{-}11}$
					& $0$                        & $6.903{\times}10^{{-}15}$   		& $1.563{\times}10^{{-}20}$
					& $\!\!0$                    & $4.431{\times}10^{{-}16}$    		& $1.768{\times}10^{{-}21}$\\
                          		&$\!\!\sigma_q{/}\mu$&$\!\!8.48{\times}10^{2}$     		& $2.637$
					& $0$                         & $5.969{\times}10^{{-}2}$     		& $8.982{\times}10^{{-}5}$
					& $\!\!0$                     & $1.515{\times}10^{{-}2}$     		& $3.026{\times}10^{{-}5}$   \\ \hline
\multicolumn{10}{p{1.900\columnwidth}}{\footnotesize {\sl Note:} Information of flow $f_{i^*}$: 1) the tuple of source, destination and demand $(s_{i^{*}}{,}t_{i^{*}}{,}d_{i^{*}}){=}(4,3,20.25)$; 2) the full SEED set $\mathcal{S}{=}\{\{2,5\},\{7,8\},$ $\{2,3,8\},\{2,8,12\},\{5,6,7\},\{5,7,11\}\}$; 3) the partial SEED set with good coverage $\mathcal{S}^\prime {=} \{\{2,5\},\{7,8\}\}$.}\vspace{5pt}\\
\end{tabular*}}\vspace{5pt}
\end{table*}

\begin{figure*}[!tbp]
    \begin{minipage}{0.30\textwidth}\centering
        \includegraphics[width=0.90\textwidth,height=0.575\textwidth]{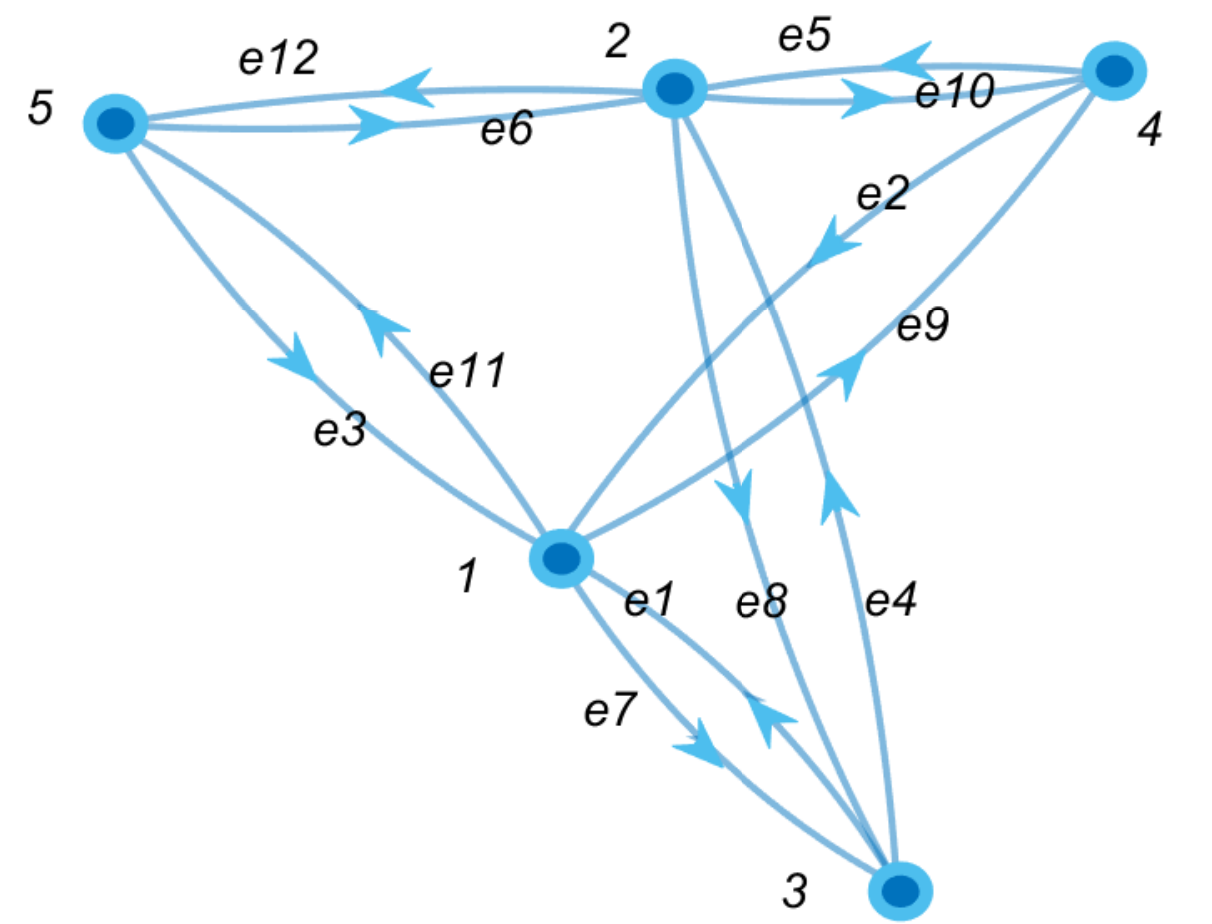}
        \vspace{5pt}
        \caption{\small Topology of an illustrative network.}\vspace{10pt}
        \label{fig:toy-case_topo}
    \end{minipage}
    %\hfill\hfil
    \begin{minipage}{0.65\textwidth}\centering
        \subfigure[CDF of coefficient of variation $\varepsilon_{\emph{CV}}$]
        {\label{fig:multflow_cv}
        \includegraphics[width=0.435\textwidth,height=0.255\textwidth]{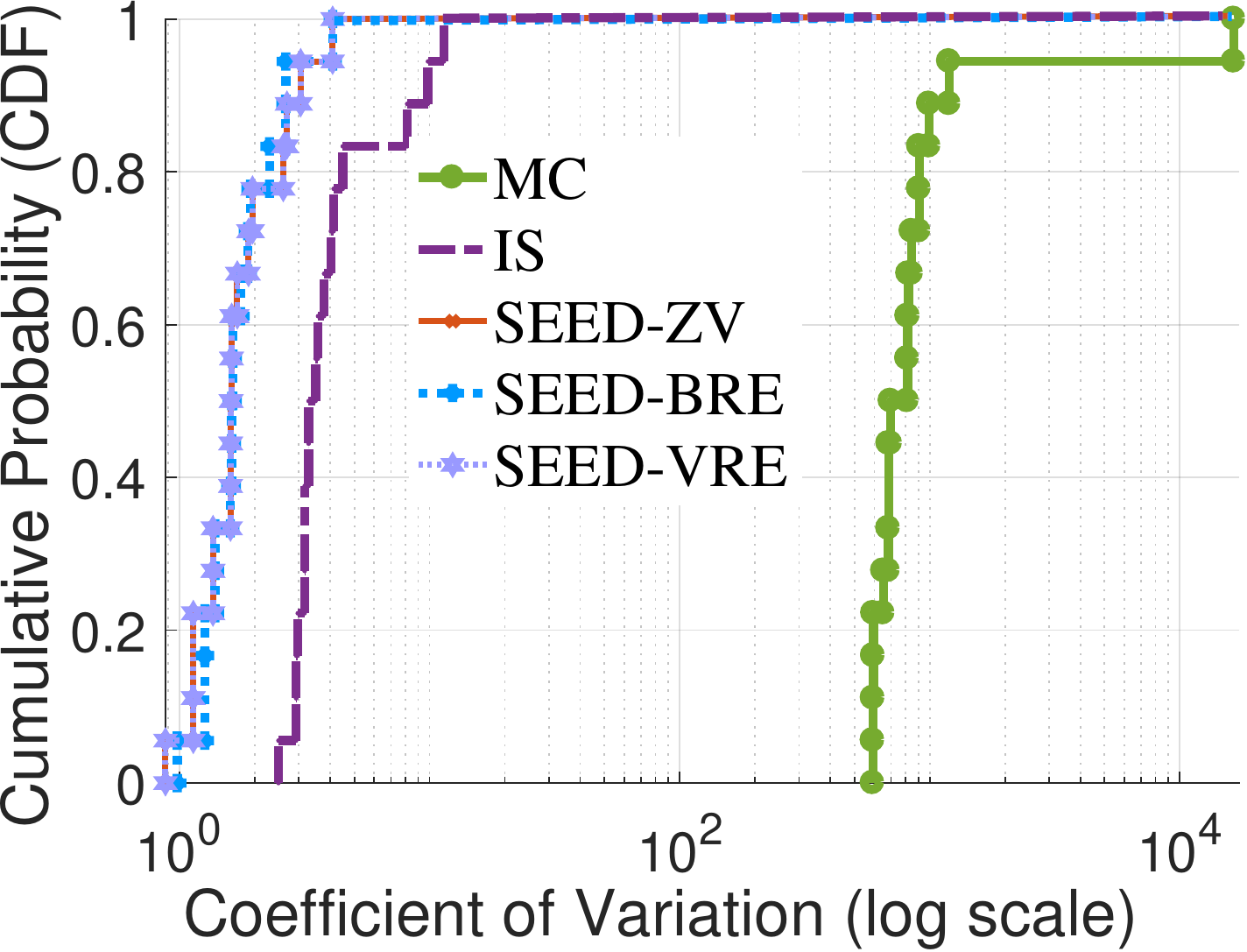}}
        \subfigure[CDF of variance reduction ($ \sigma_{\rm MC}^2{/}\sigma_q^2$)]
        {\label{fig:multiflow_vr}
        \includegraphics[width=0.435\textwidth,height=0.255\textwidth]{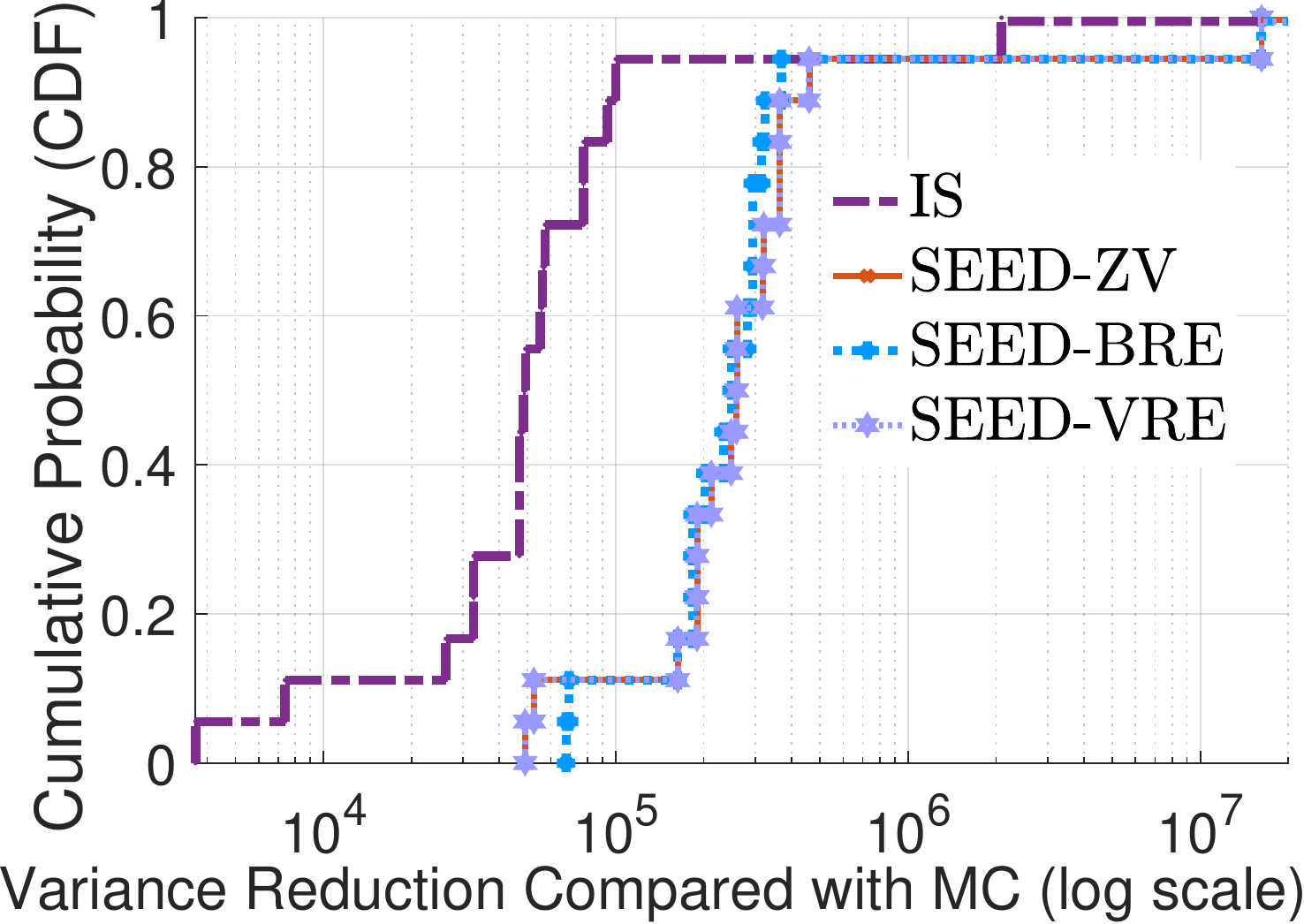}}
        \vspace{-8pt}
        \caption{\small{Performance comparison for the illustrative network (multiple flow case).}}
        \vspace{18pt}
    \end{minipage}
\end{figure*}

{\noindent \bf Experiment setting:}
The network is modelled as a directional multigraph $G$ 
with five nodes and 12 links as depicted in Fig.\ref{fig:toy-case_topo}.\newline
For each link $e_i$: 
\onlytech{the link failure probability} $p_i$ is uniformly distributed over $[0.5\epsilon,1.5\epsilon]$
($\epsilon$ is a small positive number); 
\onlytech{the link capacity} \onlypaper{$c_i$} is uniformly distributed~over $\{50,80,100,200\}$. The flow set $F$ contains 18 flows. For~each flow $f_k$: the source and destination are randomly selected; \onlytech{the demand} $d_k$ is the bandwidth demand and uniformly~distributed over $[5,25]$.
We consider traffic engineering follows 
the shortest \onlytech{\!\newline} path and max-min fairness policies. 
Note that the above setting provides 
an instance of network routing function 
$\mathcal{R}\bk{\cdot}$ in~Eq.\bk{\ref{eq:te_func_02}}.

{\noindent \bf Single flow analysis:} 
We start with the single flow case,~and\newline
select one particular flow $f_{i^*}$ from the 18 flows. 
The detailed information of $f_{i^*}$, 
together with the SEED set information,~are
introduced in notes of Table\! \ref{tb:toy-case}.
We compare our SEED methods with MC and baseline IS.~The comparison result, 
including the expectation $\mu$, 
theoretical~one-run variance $\sigma_q^2$ 
and CV $\varepsilon_{\emph{CV}}$, 
are summarized~in Table\!~\ref{tb:toy-case}. 
Let $\epsilon{=}0.05$: given a full~SEED set~$\mathcal{S}$, 
SEED-BRE and SEED-VRE achieve variance reductions~of
around {\it 2,000 and 360,000 times} compared with MC,~and 
around {\it 45 and 7,000 times}~compared with baseline IS; 
given a partial~SEED set~$\mathcal{S}^\prime$, 
our SEED methods estimate~flow~availabilities 
with very~small biases and much smaller variances,~i.e.,\!\!\newline 
with a small simulation cost, 
the estimation can be very close to the theoretical value.
We also reduce $\epsilon$ from 0.05 to 0.001, 
to validate the vanishing property of SEED methods. 
While \onlytech{the flow failure probability} $\mu$\newline reduces with the decreasing $\epsilon$: 
$\varepsilon_{\emph{CV}}$ of MC increases significantly 
as we have discussed in Section \ref{sec:pro_def_fave}; 
$\varepsilon_{\emph{CV}}$ of baseline IS is relatively stable; 
$\varepsilon_{\emph{CV}}$ of SEED methods reduces significantly, 
and $\varepsilon_{\emph{CV}}$ of SEED-VRE even achieves a {\it 300 times reduction}.

\begin{figure*}[t]
    \begin{minipage}{0.30\textwidth}\centering
        \includegraphics[width=0.90\textwidth,height=0.70\textwidth]{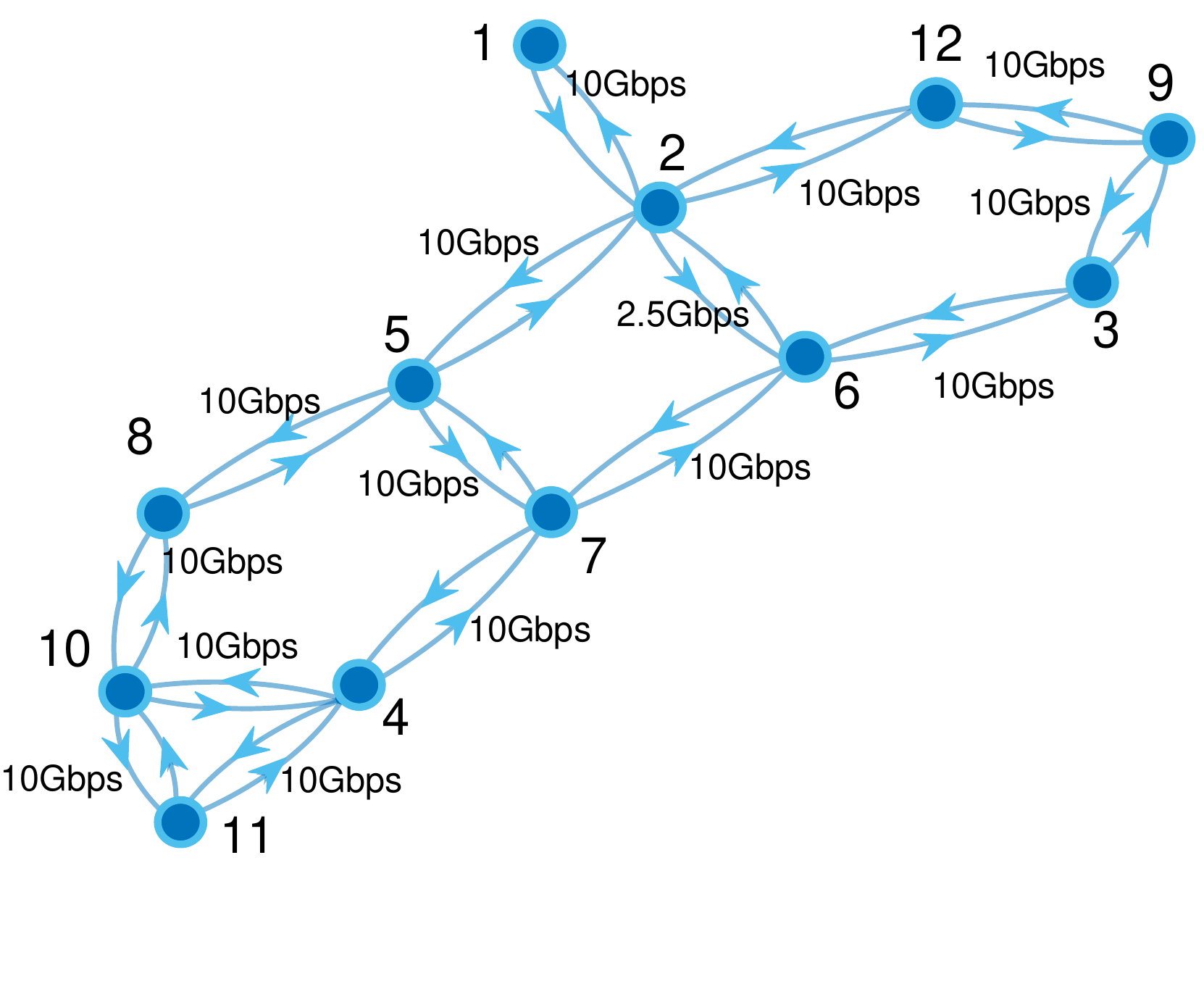}
        \vspace{-9pt}
        \caption{\small Topology of the Abilene network.}\vspace{8pt}
        \label{fig:abilene_topo}
    \end{minipage}
    %\hfill\hfil
    \begin{minipage}{0.65\textwidth}\centering
        \subfigure[Simulation cost $N$ to guarantee ``with 95\%\newline confident the relative error ${\leq}0.01$"]
        {\label{fig:realnet_sc}
        \includegraphics[width=0.435\textwidth,height=0.255\textwidth]{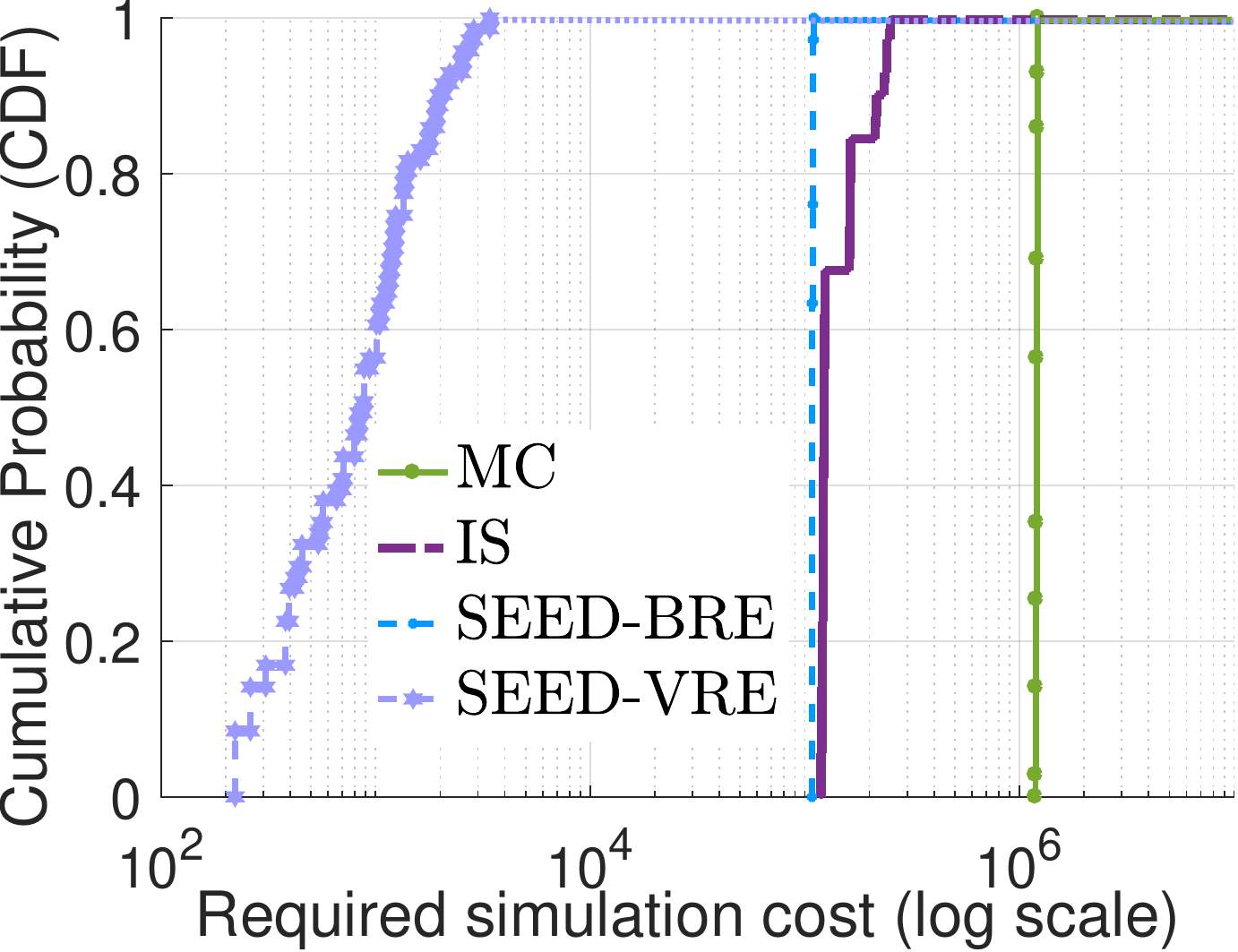}}
        \subfigure[CDF of variance reduction ($ \sigma_{\rm MC}^2{/}\sigma_q^2$)]
        {\label{fig:realnet_vr}
        \hfill
        \includegraphics[width=0.435\textwidth,height=0.255\textwidth]{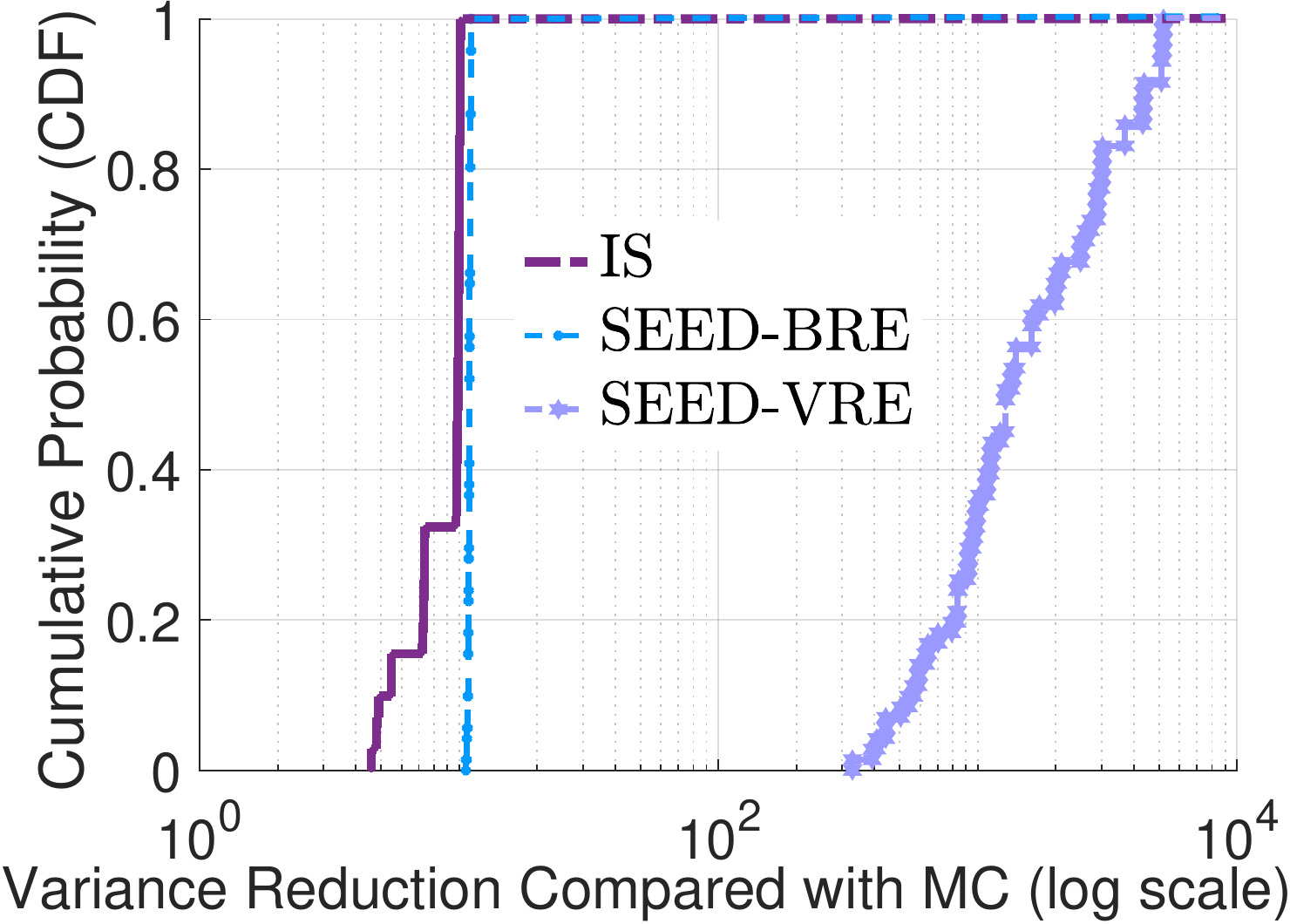}}
        \vspace{-5pt}
        \caption{\small{Performance comparison for the Abilene network (multiple flow case).}}
        \vspace{12pt}
    \end{minipage}
\end{figure*}

{\noindent \bf Multiple flows analysis}:
Next, we take all 18 flows into~consi-\newline deration. 
Let $\epsilon{=}0.001$. We consider an equally weighted~sum~of\!\newline
flows' pure importance distributions as the mixture SIS distri-bution. 
Fig.\ref{fig:multflow_cv} illustrates cumulative distributions of $\varepsilon_{\emph{CV}}$~if pure distributions~are~generated by different methods.
With SEED methods,~for around 80\% flows $\varepsilon_{\emph{CV}}{\leq}2$, 
which is smaller than the best case~of~$\varepsilon_{\emph{CV}}$ of baseline IS. 
This demonstrates~the BRE~property of SEED methods as stated in Thm. \ref{thm:mix_sis_bre}. 
Furthermore, both SEED methods and baseline IS, 
their $\varepsilon_{\emph{CV}}$ are {\it 1,000 times smaller} than that of MC. 
To depict~the variance~reduction\newline
compared with MC much clearer, 
Fig.\ref{fig:multiflow_vr} shows cumulative distributions of the variance reduction 
compared with MC\onlytech{ (in log scale)}.
\onlytech{We find that, using SEED methods,}
\onlypaper{With SEED~methods,} more than 80\% flows 
have {\it variance~reductions} $\sigma_{\emph{MC}}^2/\sigma_{\emph{SEED}}^2{>}200,000$. 
To better~illustrate the~effici-ency improvement, 
Table\!~\ref{tb:is_example} summarizes simulation costs~to guarantee that for 80\% flows, 
``with 95\%~confident the relative error is less than 0.01", 
i.e., \onlytech{the 95\% CI~width satisfies} ${\alpha_{95} \sigma_q}/{\sqrt{N}}{\leq}0.01\hat{\mu}$.

\vspace{2pt}
\begin{table}[H]
\vspace{5pt}
\small
\centering
\caption{Summary of simulation costs ($N$ steps)}
\label{tb:is_example}
\begin{tabular}{|c||c|c|c|}\hline
                        		    	& MC                   		& IS                  			& SEED-BRE\\\hline
Simulation cost ($N$)   	& $3.11\times10^{10}$  	& $7.78\times10^{5}$	& $2.01\times10^{5}$\\\hline
\end{tabular}
\end{table}\vspace{-8pt}

%%%
\subsection{\textbf{Experiments on a Realistic Network}}
\label{subsec:realnet}
Next, consider a realistic network to demonstrate 
the ``improvements in practice" by using our methods. 
As it is hard~to\newline
obtain the full SEED sets information in the complex realistic\newline
case, simulations on the realistic network can validate the~efficiency of SEED methods 
when estimating all flows' availabilities, 
given partial SEED sets with poor coverage property.

{\noindent \bf Experiment setting:}
We use the Abilene network \cite{Abilene,Jiang2009-abilene}
with topology and traffic matrices collected by \cite{Zhang2004-abilene}.
The~network contains 12 nodes and 30 links\onlytech{, and link capacities~are\newline
illustrated in Fig.\ref{fig:abilene_topo}}. 
The flow set contains~132 aggregated~flows: \onlytech{\!\newline}
all flows with the same source and destination are aggregated as 
a single flow\footnote {We take the Abilene network as an example 
and consider aggregated flows due to limitations of the accessible realistic traffic data.}. 
We take each flow's~peak\newline
(99~percentile) throughput \cite{raicu2006-harnessing} 
\onlytech{during one month} as its raw demand.
As Abilene has a sufficient capacity to serve raw demands,
we double~raw demands to see whether the network can still support oversubscribed demands. 
The routing follows the shortest path policy. 
The capacity allocation follows the max-min fairness policy, 
which is also adopted by Google's B4 backbone network \cite{jain2013-b4}.

{\noindent \bf Multiple flows analysis}: 
We consider all aggregated flows~and estimate their availabilities at the same time. 
Due to the high dimensionality of FAVE in this realistic network, 
it is costly~to obtain theoretical variances of different methods. 
Thus, we~run\newline each method 10,000 times and 
use the {\it empirical variance}~$\hat{\sigma}_q^2$ \cite{muller2016-improved} 
to estimate the one-run variance $\sigma_q^2$ and 
compute the~simulation cost $N$ to 
guarantee that~``with 95\% confident the~relative error is below 0.01". 
Fig.\ref{fig:realnet_sc} shows cumulative~distribu-tions of $N$ by 
taking the mixture of pure distributions~gener-ated by 
different methods\footnote{Due to the exponential complexity, SEED-ZV is not applied in this case.}.
With SEED-VRE, we find that~to achieve the desired accuracy level, 
for around 60\% flows the required simulation costs $N{\leq}1,000$, 
and for around~80\%~flows $N{\leq}1,400$. 
Simulation costs for SEED-BRE, baseline IS and MC methods 
to guarantee 80\% flows to achieve the accuracy target are 
100,000, 180,000 and 1,260,000, respectively. 
So the\newline efficiency is improved by 
around {\it 900~times} via SEED-VRE~and 
{\it 13 times} via SEED-BRE, 
compared with MC. 
Fig.\ref{fig:realnet_vr}~illustr-ates cumulative distributions of the variance reduction 
compared with MC\onlytech{ (in log scale)}. 
With SEED~methods, \onlytech{more than} 80\% of the flows have variance reductions
\onlytech{$\sigma_{\emph{MC}}^2/\sigma_{\emph{SEED}}^2$} larger than $900$~times.

\begin{figure*}[!tbp]\centering
        \begin{minipage}{0.420\textwidth}\centering
        \subfigure[]{\label{fig:hotlink_util}
        \includegraphics[width=0.31\textwidth,height=0.47\textwidth]{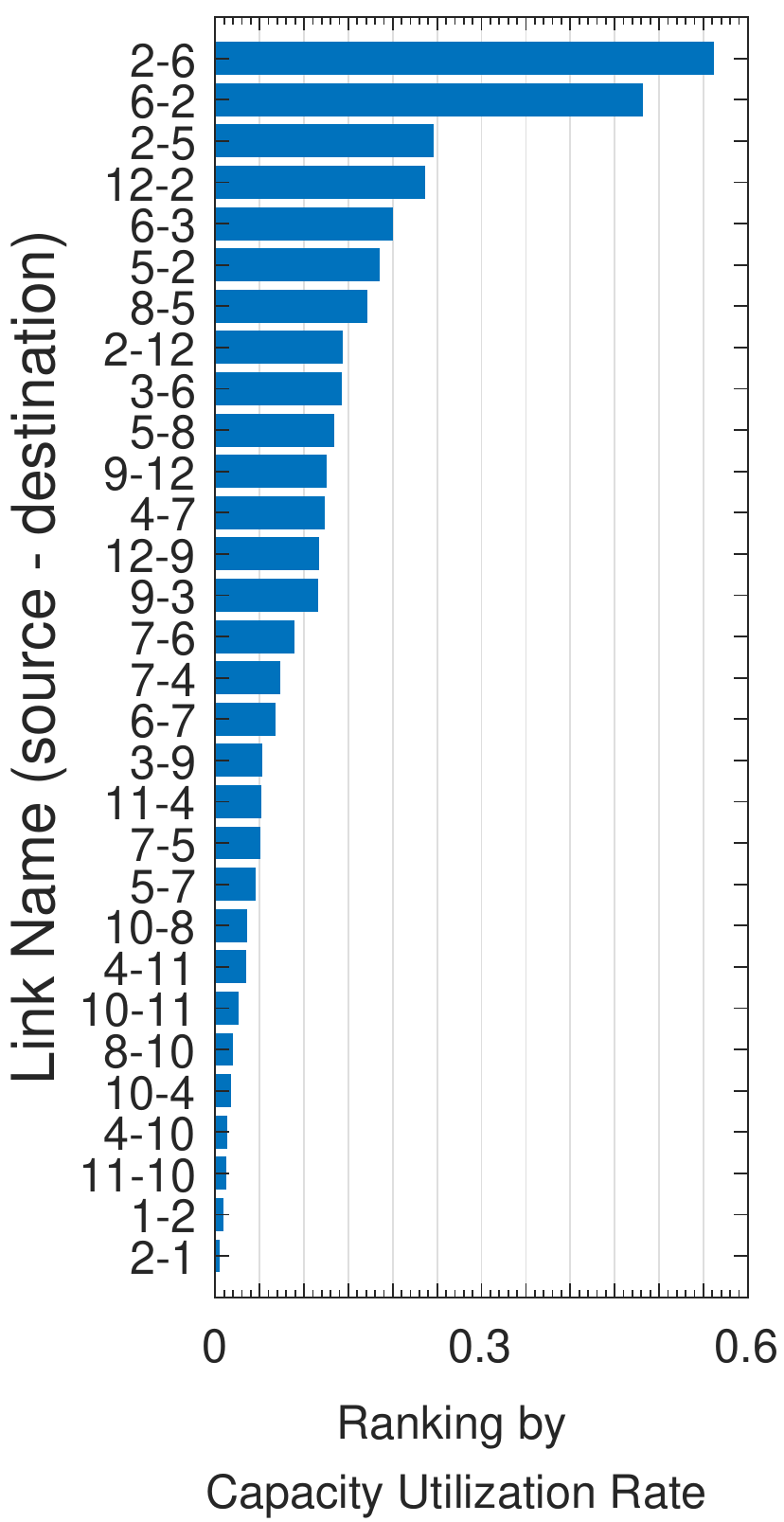}}\hfil\hfill
        \subfigure[]{\label{fig:hotlink_maxf}
        \includegraphics[width=0.31\textwidth,height=0.47\textwidth]{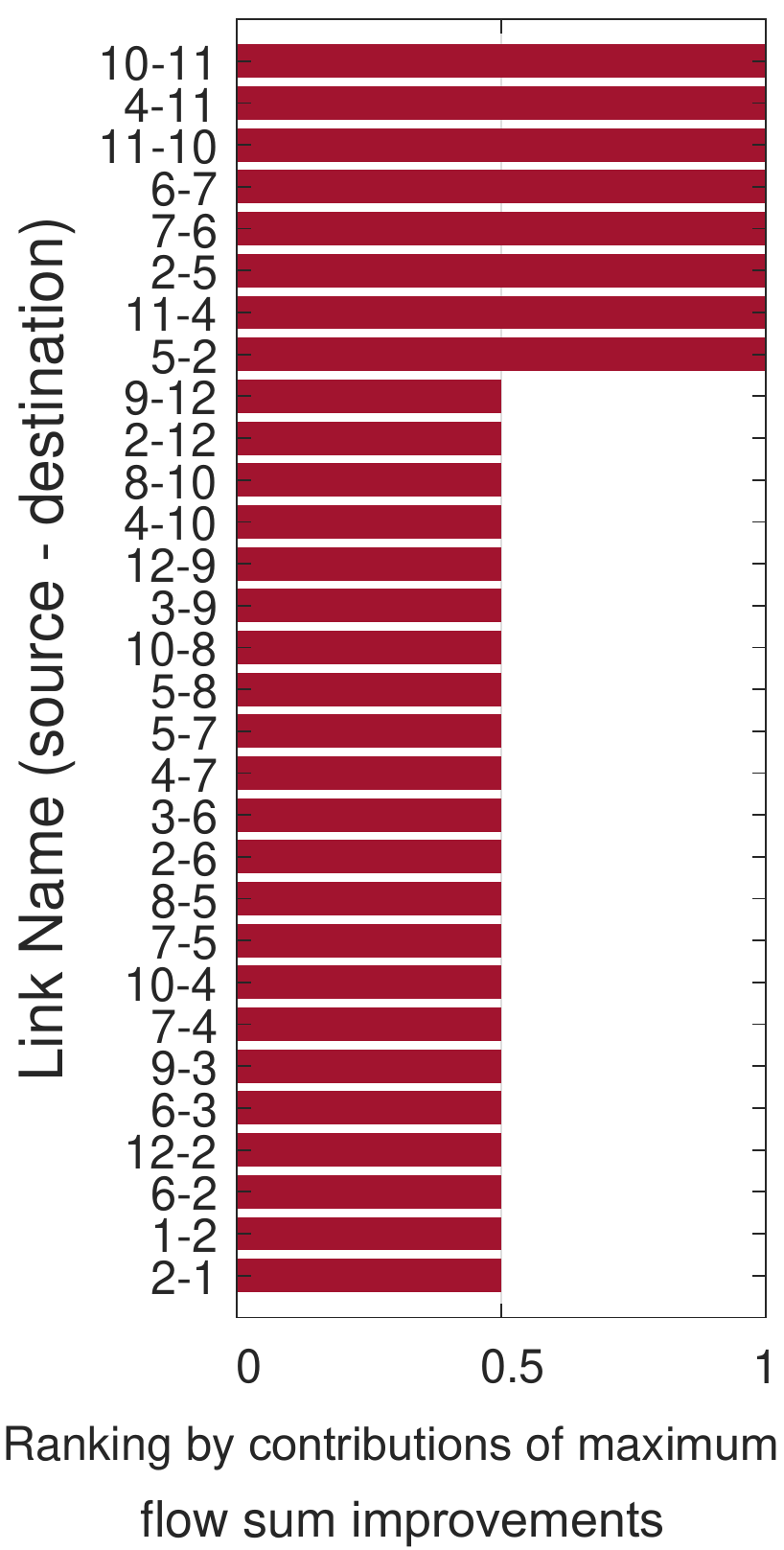}}\hfil\hfill
        \subfigure[]{\label{fig:hotlink_fave}
        \includegraphics[width=0.31\textwidth,height=0.47\textwidth]{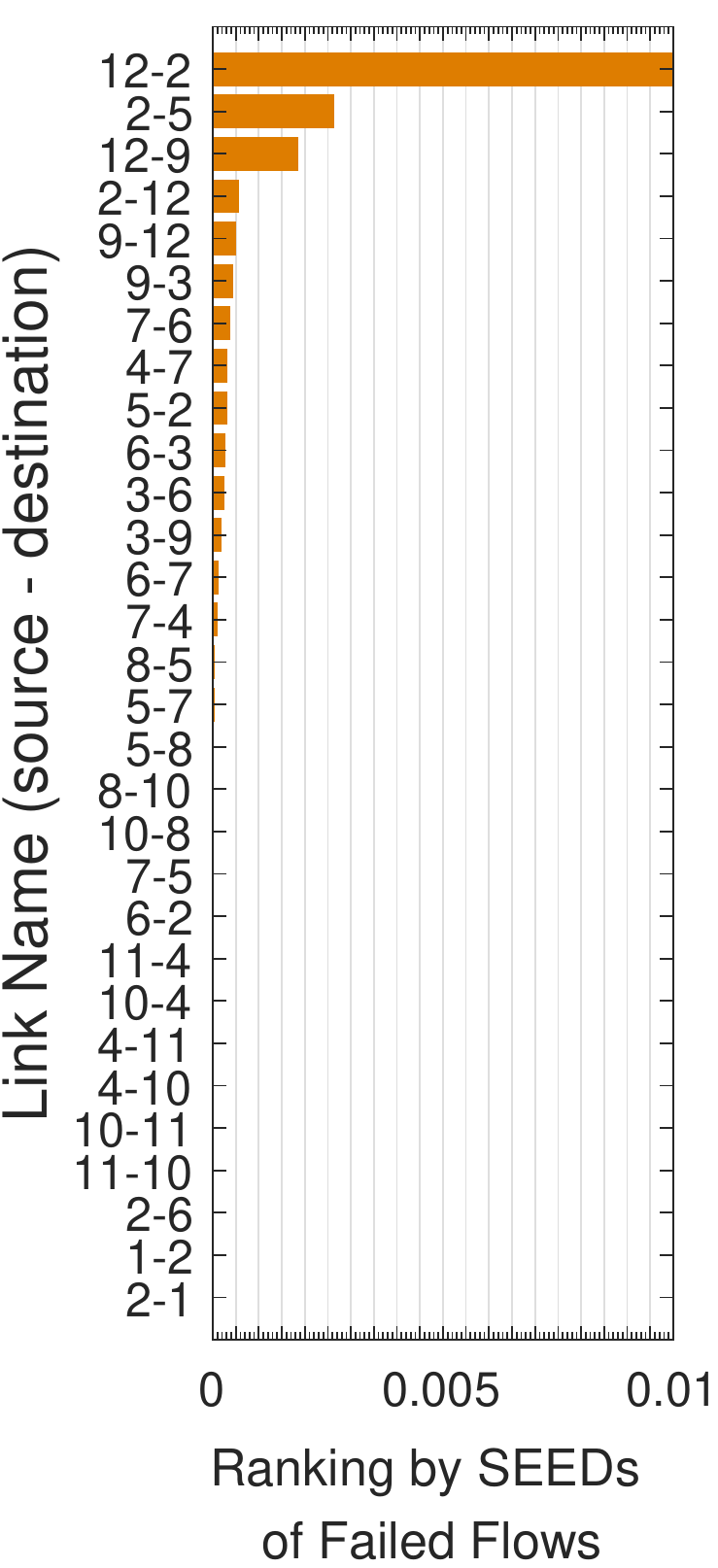}}\vspace{-8pt}
        \caption{Link importance ranking}\vspace{12pt}\label{fig:hotlink}
        \end{minipage}
	   \begin{minipage}{0.560\textwidth}\centering
	   \subfigure[CCDF of flow availability] {\label{fig:cap_fav}
        \includegraphics[width=0.480\textwidth,height=0.335\textwidth]{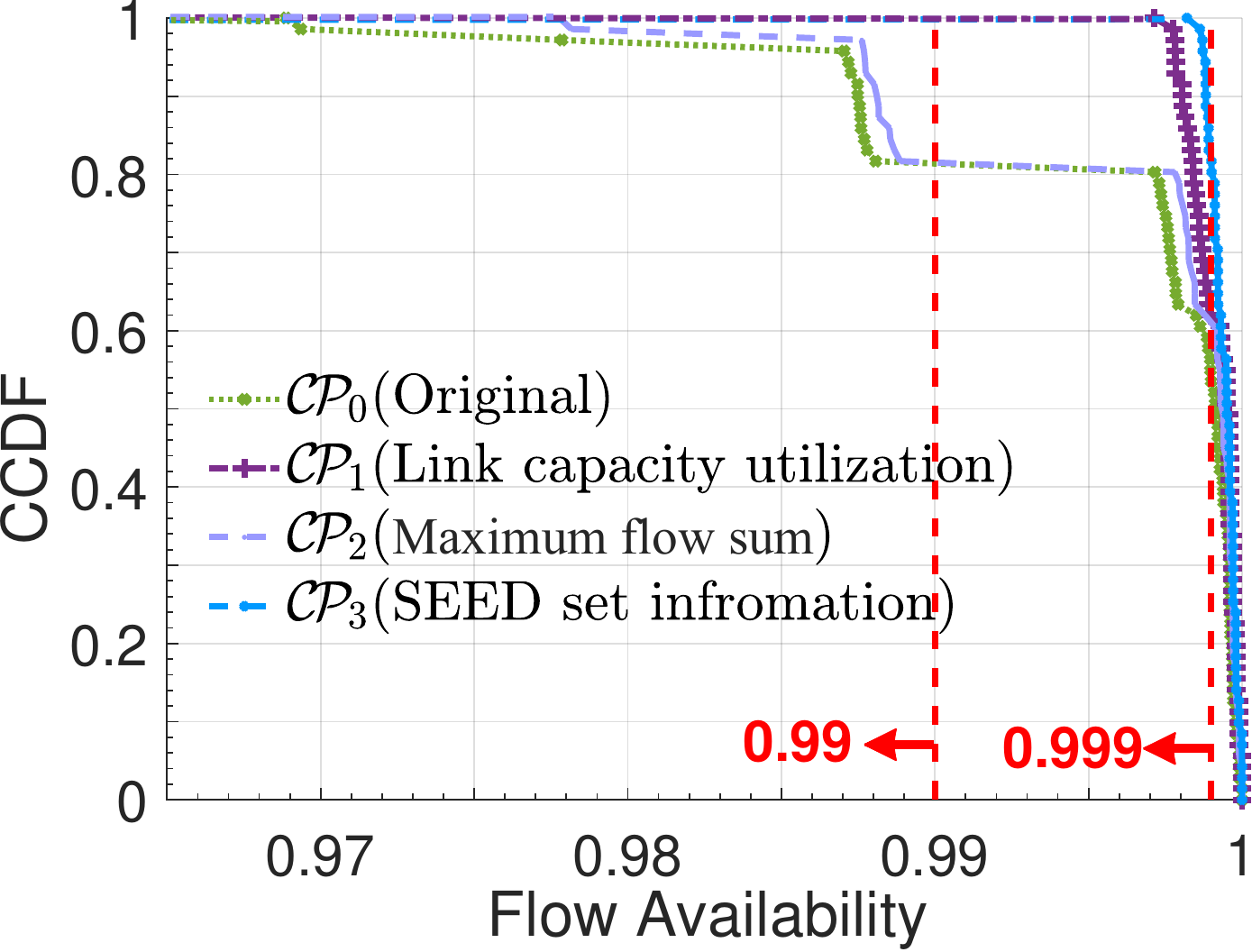}}\hfil
        \subfigure[CCDF of maximum flow] {\label{fig:cap_fde}
        \includegraphics[width=0.430\textwidth,height=0.335\textwidth]{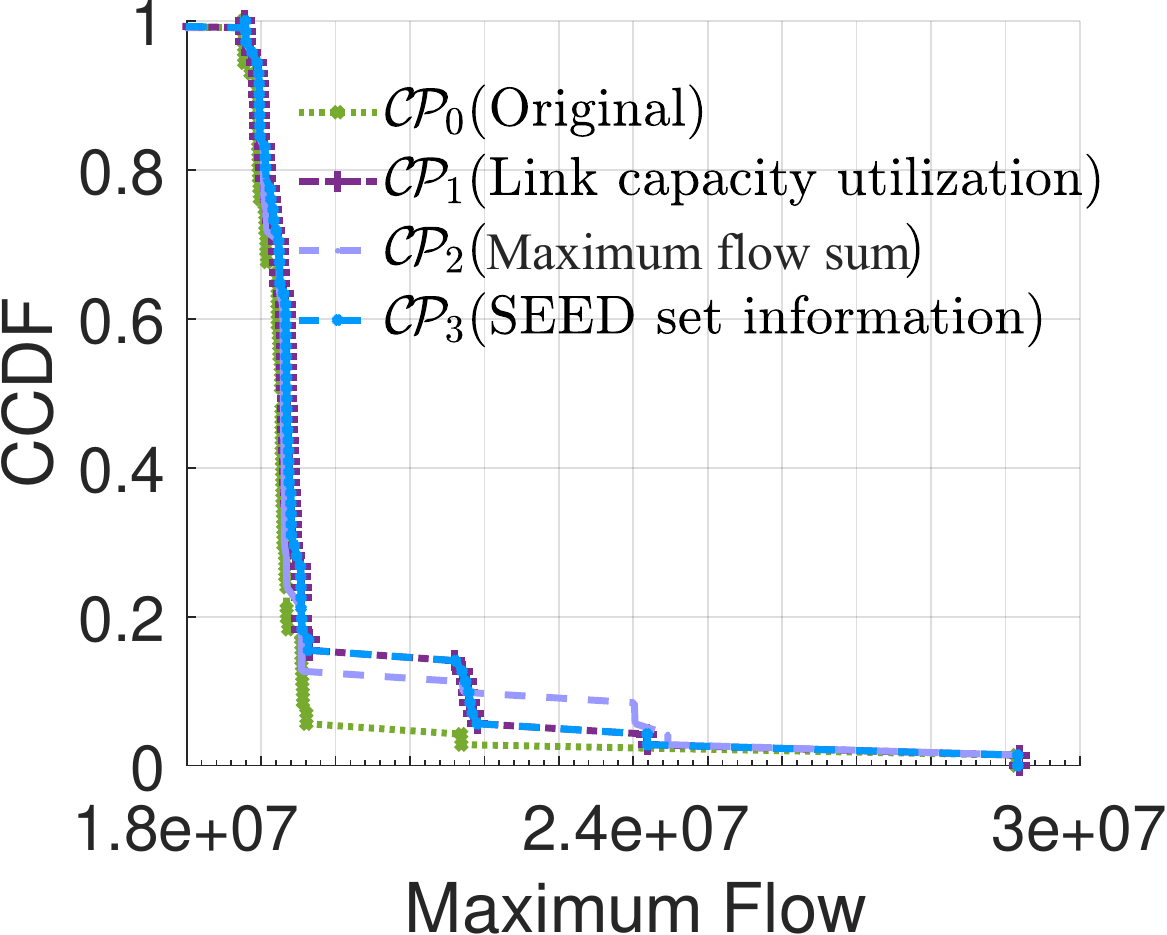}}
        \vspace{-6pt}
        \caption{An example of applying FAVE to the capacity planning.}\label{fig:capeva}\vspace{11pt}
        \end{minipage}
\end{figure*}
%%%%%%
%%%%%%
\section{\textbf{Applications in Capacity Planning}}
\label{sec:app}\vspace{-3pt}
Now, we demonstrate the utility of our method 
in capacity planning on the Abilene network. 
\onlytech{We leave the discussion for sales and topology planning in our technical report~\cite{tech}.}

Consider the case where a network provider needs to~build 
new links with certain capacities 
so as to achieve all flow~availability targets. 
For a capacity planning proposal $\mathcal{CP}$, 
we~can {\it evaluate its feasibility and adopt it if feasible}, as follows:
\begin{itemize}[leftmargin=3mm]
\item 
We obtain {\it availability feedbacks}~of each flow $f_i$, 
including a flow availability estimation $\hat{\mu}_i$~with 
upper (lower) confidence bound $\textstyle\overline{\mu_i}$ ($\textstyle\underline{\mu_i}$) 
computed via~empirical variances.\vspace{3pt}
\item 
$f_i$'s availability target $o_i$ is {\it achieved} 
if $\textstyle\underline{\mu_i}{\geq}o_i$ and 
{\it unreached} if $\textstyle\overline{\mu_i}{<}o_i$.\!\footnote{We run enough simulations to guarantee $o_i{\not\in}[\underline{\mu_i},\overline{\mu_i}]$.} 
$\mathcal{C\!P}$ is {\it feasible} if all flows' availability targets are achieved; 
and infeasible otherwise.
\end{itemize}
By evaluating flow availabilities, 
we can not only {\it determine which proposal allows the network to 
provide better~flow availability guarantees}, 
but also {\it utilize flow availability feedbacks 
for further~refinements of infeasible proposals}. 
We consider~the following information to refine an infeasible~proposal:\!\!
\begin{itemize}[leftmargin=3mm]
\item 
{\bf Link capacity utilization based}: 
The utilization metric~is~the\!\!\newline primary metric of interest 
in capacity planning~\cite{Cisco-MATE2}. 
Hence, link capacity utilizations can imply the importance of links.
\item 
{\bf Maximum flow based}: 
The maximum flow value is a~widely\!\newline adopted 
network reliability measure~\cite{botev2014reliability,gertsbakh2014permutational,datta2016reliability}. 
We take the increase of the sum of maximum flows 
brought by increasing one unit link capacity to measure the importance of links.
\item 
{\bf SEED based:} 
By testing flow availabilities, 
we get $\mathcal{F}^\prime$,~the set of flows with unsatisfied availability targets. 
The~number\newline of failed flows in $\mathcal{F}^\prime$ 
when $e_j$ fails ($\sum_{f_i\!{\in}\mathcal{F}^\prime}\mathbb{P}\left[\mathcal{R}_i{=}1|x_j{=}1\right]$) 
can be estimated using SEED algorithms 
and can imply the importance of links.\footnote{As $\mathbb{P}[\mathcal{R}_i{=}1|\pmb{x}_{1{:}k}]$, the probability that flow $f_i$ fails given the first $k$th links' statues $\pmb{x}_{\!1{:}k}$, can be estimated using SEED algorithms, $\mathbb{P}\left[\mathcal{R}_i{=}1|x_j{=}1\right]$ can also be estimated by taking $e_j$ as the first link and let $k{=}1$, $\pmb{x}_{1:k}{=}1$.}
\end{itemize}

Let the current capacity design of Abilene be the original proposal $\mathcal{CP}_{\!0}$, 
and twice the raw demands be flow demands, 
such that $\mathcal{CP}_{\!0}$ is infeasible. 
We rank links using the above~metrics and summarize rankings in Fig.\ref{fig:hotlink}. 
Assume the provider has a budget and only affords to build four new links, 
each has a capacity of 2.5Gbps and a failure probability of 0.01. 
Based~on rankings in Fig.\ref{fig:hotlink},~we~have three proposals, 
i.e., $\mathcal{CP}_{\!1}$, $\mathcal{CP}_{\!2}$ and $\mathcal{CP}_{\!3}$ 
by taking the top four links in Fig.\ref{fig:hotlink_util}, 
\ref{fig:hotlink_maxf} and \ref{fig:hotlink_fave}. 
Fig.\ref{fig:cap_fav} shows flow~availability evaluation results. 
As our method selects links with the largest impact on flow failures, 
it achieves {\it greater flow availability improvements}: in $\mathcal{CP}_{\!3}$, 
flow availabilities of around 80\% of the flows reach 99.9\%.

\noindent{\bf Insights 1:} {\it Improper capacity planning offers little help~on improving flow availabilities}. E.g., although $\mathcal{C\!P}_{\!2}$ maximizes the sum of maximum flows, it does not consider the distribution~of traffic demands across the network, and thus only brings little improvements on flow availabilities.

\noindent{\bf Insights 2:} {\it Link utilization is not always the best indicator~of capacity planning}. With the shortest path routing policy, link~$e_i$\!\newline has a high capacity utilization if many flows' shortest paths~go through it. Yet, if it is easy to find some alternate link~when~$e_i$ fails, $e_i$'s failure will not result in flow failures and so $e_i$~is~not\newline the most important if aiming at improving flow~availabilities.

\onlytech{
\begin{itemize}[leftmargin=3mm]
\item \textbf{Improper capacity planning offers little help on improving flow availabilities}. For instance, although $\mathcal{C\!P}_{\!2}$ can~maximize the sum of maximum flows, it does not consider the distribution of traffic demands across the network, and thus only brings little improvements for flow availabilities.
\item {\bf Link utilization may not always be the best indicator of capacity planning}.
When taking the shortest path routing policy, link $e_i$ will have a high capacity utilization if many flows' shortest paths go through it. However, if we can easily find some alternate links when $e_i$ fails, then $e_i$'s failure will not result in flows' failures and so $e_i$ is not that important when aiming at improving flow availabilities.
\end{itemize}}

\onlytech{As our method select links which have the largest~impact~on\newline flow failures, it achieves {\it greater flow availability improvements\!\newline when refining capacity proposals}: in $\mathcal{CP}_{\!3}$, flow availabilities of around 80\% of the flows reach 99.9\%.
Our method also provides {\it more accurate evaluations~to~select better capacity proposals}.
According to Fig.\ref{fig:cap_fav},~we can easily determine a better proposal by comparing $\mathcal{CP}_{\!1}$, $\mathcal{CP}_{\!2}$~and $\mathcal{CP}_{\!3}$ over flow availability. However, as illustrated~by~Fig.\ref{fig:cap_fde}, it is hard to select a better proposal by comparing these three proposals over the sum of maximum flows (traditional~method).}

\section{\textbf{Conclusion}}\label{sec:conclusion}
In this paper, we propose fast and accurate methods in~solving the FAVE problem. We introduce the concept of ``SEED"~to\newline
determine the importance of roles played by different links~in flow failures, and propose three SEED based SIS methods which achieve BRE and VRE properties with linear~computational complexities. To provide robust and scalable estimations, we extend FAVE to the multiple flows case and partial SEED set case, and our methods maintain the estimation~efficiency. We apply our methods on both an illustrative network and a realistic network, and our methods reduce the simulation cost by around 900 and 130 times compared with MC and baseline IS methods on the Abilene network. We show that our method facilitates capacity planning by providing~more~accurate network reliability estimations compared with classical methods, and greater flow availability improvements compared with~solely using link capacity utilizations for capacity planning.

%%%
\bibliographystyle{IEEEtran}
\bibliography{FAVE}

% Generated by IEEEtran.bst, version: 1.14 (2015/08/26)
\begin{thebibliography}{10}
\providecommand{\url}[1]{#1}
\csname url@samestyle\endcsname
\providecommand{\newblock}{\relax}
\providecommand{\bibinfo}[2]{#2}
\providecommand{\BIBentrySTDinterwordspacing}{\spaceskip=0pt\relax}
\providecommand{\BIBentryALTinterwordstretchfactor}{4}
\providecommand{\BIBentryALTinterwordspacing}{\spaceskip=\fontdimen2\font plus
\BIBentryALTinterwordstretchfactor\fontdimen3\font minus
  \fontdimen4\font\relax}
\providecommand{\BIBforeignlanguage}[2]{{%
\expandafter\ifx\csname l@#1\endcsname\relax
\typeout{** WARNING: IEEEtran.bst: No hyphenation pattern has been}%
\typeout{** loaded for the language `#1'. Using the pattern for}%
\typeout{** the default language instead.}%
\else
\language=\csname l@#1\endcsname
\fi
#2}}
\providecommand{\BIBdecl}{\relax}
\BIBdecl

\bibitem{Cisco-MATE2}
Cisco.Com, ``Best practices in core network capacity planning,''
  \url{https://communities.cisco.com/docs/DOC-35673}, 2013.

\bibitem{Cisco-MATE}
------, ``Planning and designing networks with the {Cisco} {MATE} portfolio,''
  \url{https://communities.cisco.com/docs/DOC-36973}, 2013.

\bibitem{Facebook-Prophet}
S.~J. Taylor and B.~Letham, ``Prophet: forecasting at scale,''
  \url{https://research.fb.com/prophet-forecasting-at-scale/}, 2017.

\bibitem{Google-WAND}
A.~K. Bangla, A.~Ghaffarkhah \emph{et~al.}, ``Capacity planning for the
  {Google} backbone network,'' in \emph{Proc. ISMP}, 2015.

\bibitem{hammersley1964-MC_def}
J.~M. Hammersley and D.~C. Handscomb, ``The general nature of {Monte Carlo}
  methods,'' in \emph{Monte Carlo Methods}, 1964, pp. 1--9.

\bibitem{Glynn1989-IS_def}
P.~W. Glynn and D.~L. Iglehart, ``Importance sampling for stochastic
  simulations,'' \emph{Manag. Sci.}, 1989.

\bibitem{jiang2012-reliability_mc}
Y.~Jiang, R.~Li \emph{et~al.}, ``The method of network reliability and
  availability simulation based on {Monte Carlo},'' in \emph{Proc. IEEE
  ICQR2MSE}, 2012.

\bibitem{Yeh2010-PSO_MC}
W.~C. Yeh, Y.~C. Lin \emph{et~al.}, ``A particle swarm optimization approach
  based on {Monte Carlo} simulation for solving the complex network reliability
  problem,'' \emph{IEEE Trans. Rel.}, 2010.

\bibitem{Ecuyer2011-ZV_SIS}
P.~L'Ecuyer, G.~Rubino \emph{et~al.}, ``Approximate zero-variance importance
  sampling for static network reliability estimation,'' \emph{IEEE Trans.
  Rel.}, 2011.

\bibitem{Lee2010-crosslayer_reliability}
K.~Lee, H.-W. Lee, and E.~Modiano, ``Reliability in layered networks with
  random link failures,'' in \emph{Proc. IEEE INFOCOM}, 2010.

\bibitem{botev2014reliability}
Z.~I. Botev, S.~Vaisman \emph{et~al.}, ``Reliability of stochastic flow
  networks with continuous link capacities,'' in \emph{Proc. IEEE WSC}, 2014.

\bibitem{gertsbakh2014permutational}
I.~Gertsbakh, R.~Rubinstein \emph{et~al.}, ``Permutational methods for
  performance analysis of stochastic flow networks,'' \emph{Probab. Eng.
  Informational Sci.}, 2014.

\bibitem{datta2016reliability}
E.~Datta and N.~K. Goyal, ``Reliability estimation of stochastic flow~networks
  using pre-ordered minimal cuts,'' in \emph{Proc. IEEE MicroCom}, 2016.

\bibitem{Abilene}
``Abilene,'' \url{https://www.internet2.edu/}.

\bibitem{xiong2005novel}
J.~Xiong and W.~Gong, ``A novel algorithm on network reliability estimation,''
  \emph{Mathematical and computer modelling}, 2005.

\bibitem{ball1986-computational}
M.~O. Ball, ``Computational complexity of network reliability analysis: An
  overview,'' \emph{IEEE Trans. Rel.}, 1986.

\bibitem{connection_availability}
A.~M. Shooman, ``Algorithms for network reliability and connection availability
  analysis,'' in \emph{Proc. IEEE ELECTRO}, 1995.

\bibitem{service_availability}
A.~E. Conway, ``Fast simulation of service availability in mesh networks with
  dynamic path restoration,'' \emph{IEEE TON}, 2011.

\bibitem{tech}
``Technical report,'' \newline
  https://dl.dropboxusercontent.com/s/5tjzn90wsys81h1/FAVE.pdf?dl=0.

\bibitem{Jiang2009-abilene}
W.~Jiang, R.~Zhang-Shen \emph{et~al.}, ``Cooperative content distribution and
  traffic engineering in {ISP} network,'' in \emph{Proc. ACM SIGMETRICS}, 2009.

\bibitem{Zhang2004-abilene}
Y.~Zhang, ``Six months of {Abilene} traffic matrices,''
  \url{http://www.cs.utexas.edu/~yzhang/}, 2014.

\bibitem{raicu2006-harnessing}
I.~Raicu, I.~Foster, and A.~Szalay, ``Harnessing grid resources to enable the
  dynamic analysis of large astronomy datasets,'' in \emph{Proc. IEEE
  Supercomputing}, 2006.

\bibitem{jain2013-b4}
S.~Jain, A.~Kumar \emph{et~al.}, ``B4: Experience with a globally-deployed
  software defined {WAN},'' in \emph{Proc. ACM SIGCOMM}, 2013.

\bibitem{muller2016-improved}
A.~M{\"u}ller, ``Improved variance reduced {Monte-Carlo} simulation of
  in-the-money options,'' \emph{J. Math. Finance}, 2016.

\end{thebibliography}

%%%
%\clearpage
%\input{sections/appendix}

\end{document}